\documentclass[11pt]{article}
\usepackage[preprint]{acl}
\usepackage{times}
\usepackage{latexsym}
\usepackage[T1]{fontenc}
\usepackage[utf8]{inputenc}
\usepackage{microtype}
\usepackage{inconsolata}
\usepackage{graphicx}
\usepackage{booktabs}
\usepackage{amsmath}
\usepackage{multirow}
\usepackage{colortbl}

\title{SkillScreen: High-Recall Screening of Malicious Agent Skills\\with Decision Models}

\author{
Jerrold Huang\textsuperscript{1*} \quad
Junwei Quan\textsuperscript{1*} \quad
Saiyue Lyu\textsuperscript{2} \quad
Zihan Guo\textsuperscript{1} \quad
Yingqiang Ge\textsuperscript{3} \\
\textsuperscript{1}University of Toronto \quad
\textsuperscript{2}University of British Columbia \quad
\textsuperscript{3}Rutgers University
}

\begin{document}
\maketitle
\begingroup\renewcommand\thefootnote{*}\footnotetext{Equal contribution.}\endgroup

\begin{abstract}
Agent skills turn natural-language instructions into actions, creating a security challenge: legitimate automation and malicious behaviour can look deceptively similar. A practical screen must catch nearly every attack while keeping the review load manageable. We present SkillScreen, a compact decision model trained to score skill maliciousness directly from \texttt{SKILL.md} in a single forward pass. On MaliciousSkillBench, SkillScreen flags 20--25\% of benign skills at 99\% recall on unseen attack families, substantially fewer than the word-level classifier and prompted LLM judges tested. Under the same training protocol, it also generalizes better than a fully fine-tuned encoder to unseen sources. Yet unfamiliar benign sources sharply increase false positives, showing that recognizing new attacks does not ensure reliable screening of new skill collections. Our results demonstrate the value of task-trained decision models for high-recall skill screening. Project page: \url{https://jerroldh.github.io/SkillScreen-webpage/}
\end{abstract}

\begin{figure*}[t]
\centering
\includegraphics[width=\textwidth]{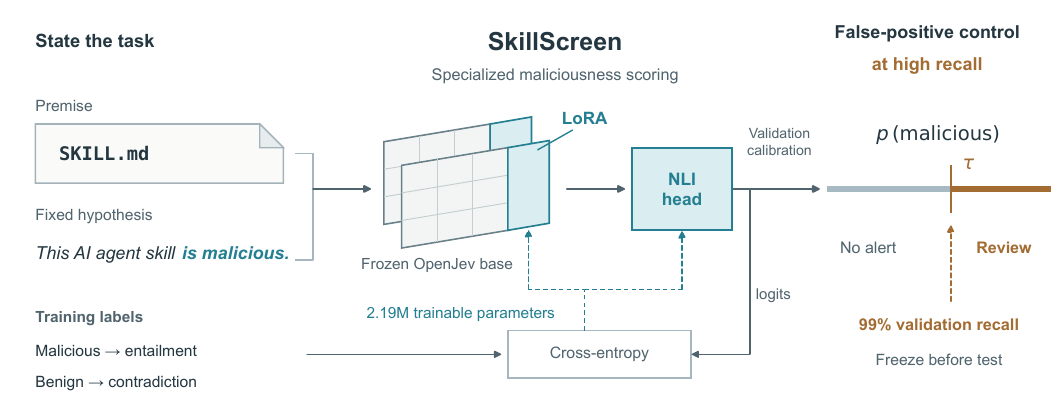}
\caption{SkillScreen specializes a small language model for false-positive control at high recall. Training and inference use the same skill, hypothesis and model. Cross-entropy supervises NLI outputs with skill labels. Dashed teal arrows mark updates to LoRA adapters and the head. Validation data calibrate scores and select a threshold for 99\% recall. Model, calibration and threshold are frozen before test scoring. Layers and score axis are schematic.}
\label{fig:overview}
\end{figure*}

\section{Introduction}

Agent skills package natural-language instructions in a \texttt{SKILL.md} file, often with scripts and configuration, that an AI agent loads to manage files, write code or query a service \citep{agentskills2026spec}. The agent carries out these instructions with the tools and permissions it has been granted, so a skill turns text into actions. Skills shared through public registries form a software supply chain. Registry analyses have identified malicious skills that steal credentials, execute remote code and manipulate agent behaviour \citep{liu2026domentionuserdetecting}. Controlled evaluations also show that skill-file attacks can induce data exfiltration and destructive actions \citep{schmotz2026skillinjectmeasuringagentvulnerability}. Registries and users therefore need to screen skills before they are installed.

Screening is hard because legitimate automation and malicious behaviour can look deceptively similar. Benign developer tools also run shell commands, read credentials and contact remote services \citep{chen2026skillleakbench}. What separates an attack is how these capabilities are used in context \citep{schmotz2026skillinjectmeasuringagentvulnerability}. A practical screen must maintain high recall, since one missed attack can compromise its users, while keeping the false-positive rate (FPR) low enough to limit human review. These goals are difficult to meet together because lowering the decision threshold to catch more attacks might also flag more benign skills.

Existing screens struggle to meet both requirements. A static scanner matches risky patterns and detects 10.5\% of malicious skills on MaliciousSkillBench \citep{skillspector2026, wang2026maliciousskillbench}. 
Prompted LLM judges can inspect skill instructions, but they are expensive to query and tend to make overly conservative predictions. Moreover, their coarse-grained scores make it difficult to achieve high recall without excessive false positives: on the benchmark's main split, all tested judges incorrectly flag at least 87\% of benign skills when operating at 99\% recall.
% Prompted LLM judges read the instructions, but they are costly per query, conservative at their default decision, and their coarse probabilities cannot be tuned to high recall: on the benchmark's main split, every judge we test flags about 87\% or more of benign skills to reach 99\% recall.

To address these limitations, we introduce SkillScreen, a lightweight decision model that detects malicious agent skills directly from \texttt{SKILL.md} in a single forward pass. As illustrated in Figure~\ref{fig:overview}, SkillScreen formulates detection as natural language inference (NLI), pairing each skill's instructions with the fixed hypothesis \emph{This AI agent skill is malicious}. By fine-tuning OpenJev-2B \citep{openjev2026} with LoRA, SkillScreen learns to produce continuous maliciousness scores that enable flexible threshold calibration. 
These scores allow us to select a threshold targeting 99\% recall on validation data. In our evaluation, SkillScreen yields fewer false positives than prompted LLM judges at matched high recall and runs locally without potential costly per-query API fees.
% This allows us to target 99\% recall using a validation-calibrated threshold while reducing false positives compared with prompted LLM judges, at substantially lower inference cost.
% These scores support validation-calibrated thresholding targeting 99\% recall, yielding fewer false positives than prompted LLM judges in our evaluation while avoiding potentially costly per-query inference API fees through local deployment.
% To tackle these challenges, we present SkillScreen, a compact decision model trained to score skill maliciousness directly from \texttt{SKILL.md} in a single forward pass. SkillScreen formulates detection as natural language inference (NLI), pairing a skill's text with the fixed hypothesis \emph{This AI agent skill is malicious}. We adapt OpenJev-2B \citep{openjev2026} with LoRA to score how strongly the text supports this hypothesis. The continuous score lets us select a review threshold targeting 99\% recall on validation data and freeze it before testing (Figure~\ref{fig:overview}).

On MaliciousSkillBench's 9,740 labelled skills from eleven sources, SkillScreen detects 95.6\% of malicious skills from attack families unseen in training at a 3.3\% FPR. It matches a fully fine-tuned ModernBERT-large encoder in this setting, and at 99\% recall it flags about half as many benign skills as a word-level SVM and under a third as many as prompted judges. Under the same one-epoch protocol, it generalizes better than the encoder to unseen sources (FPR 24 points lower at 99\% recall). Yet unfamiliar benign sources sharply increase false positives for every trained detector: recognizing new attacks does not ensure reliable screening of new skill collections. We make three contributions.
\begin{itemize}\setlength\itemsep{0pt}
\item \textbf{A compact high-recall screen.} We introduce SkillScreen with a frozen, seed-aware protocol that compares detectors at matched high recall with paired resampling, and release the model, the scripts and the per-skill predictions behind Table~\ref{tab:main}.
\item \textbf{Evidence under attack and source shift.} Across structural, source and random splits, SkillScreen keeps recall high and needs the fewest false positives at 99\% recall among the detectors we test.
\item \textbf{Mechanism and bottleneck.} Task training drives the gain, and an NLI initialization with a task-naming hypothesis halves false positives when labels are limited. Benign-source coverage, rather than attack novelty, is the central obstacle to deployment.
\end{itemize}

\section{Preliminaries}

\paragraph{Agent skills.}
The front matter of a skill specifies its name and describes when it should be used. Its body provides instructions and may include commands or code. Scripts and references can be loaded while the agent works \citep{agentskills2026spec}. Skills complement tool interfaces such as MCP \citep{mcp2025spec} by describing how the agent should use its capabilities.

We consider an attacker who publishes a skill for users to install. SkillScreen targets attacks that the skill's Markdown expresses, directly or through an initialization step it prescribes; behaviour that lives only in bundled code calls for package-level analysis (see Limitations). The detector scores the Markdown before the skill is used, and a high score routes the skill to human review.

\paragraph{Benchmark.}
MaliciousSkillBench combines marketplace collections and red-team benchmarks \citep{wang2026maliciousskillbench}. It groups malicious skills into 4,575 structural families. The Malicious-Structural-Disjoint (MSD) protocol keeps each family within a single partition. The Source-Disjoint (SD) protocol holds out entire sources for testing. The Source-Balanced Random (SBR) protocol draws training and test data from the same pool of sources. We exclude the five records whose exact public text is unavailable and keep the official labels and splits; SD's split omits eight more.

\section{SkillScreen}
\label{sec:method}
SkillScreen is a decision model that maps skill text and a fixed maliciousness hypothesis to a continuous score without generating text. A threshold chosen on the validation set routes skills to review. We initialize SkillScreen from a pinned OpenJev-2B checkpoint \citep{openjev2026}, which adapts Qwen3.5-2B \citep{qwen2026qwen35} for NLI. Figure~\ref{fig:overview} shows how task training, the NLI scoring interface and the review decision connect.

\subsection{LoRA fine-tuning}
LoRA \citep{hu2021lora} specializes the model by learning a small set of low-rank weight updates. We freeze the backbone, train rank-8 adapters with scaling factor $\alpha=16$ on the attention projections and fine-tune the classification head, initialized from its pretrained NLI weights. The adapters and head together contain 2.19M trainable parameters and can be shipped separately from the backbone.

We train for a single epoch on the labelled training partition, on one 10\,GB GPU slice. Appendix~\ref{app:implementation} details optimization, memory use and checkpoint selection.

\subsection{NLI formulation}
The NLI interface provides a pretrained classification head and states the task as an explicit proposition. In NLI, a model classifies the relation between a premise and a hypothesis \citep{yin2019benchmarking, wang2021entailmentfewshotlearner}. For each skill $x$, its Markdown text serves as the premise, paired with the following fixed hypothesis:
\begin{quote}
\emph{This AI agent skill is malicious.}
\end{quote}
The input sequence is the premise followed by the hypothesis. Inputs are capped at 8,192 tokens; for longer inputs, only the premise is truncated, so the hypothesis is always preserved in full.

The pretrained OpenJev head outputs three logits for the contradiction, entailment and neutral classes. We train with a cross-entropy loss, using contradiction as the target class for benign skills and entailment for malicious skills; no training example targets the neutral class. The same input format and head also let the untuned checkpoint score skills directly, which gives a zero-shot reference point. Stating the task in the hypothesis pays off at 99\% recall (Section~\ref{sec:recipe}), and with 25--50\% of the labels the NLI initialization and the task-naming hypothesis together halve the false-positive rate (Section~\ref{sec:dataeff}); with all labels, the recipe is equally effective from a plain Qwen3.5-2B checkpoint (Section~\ref{sec:recipe}).

\subsection{Scoring and review}
Let $z_c$, $z_e$ and $z_n$ denote the contradiction, entailment and neutral logits, respectively. The detection score is defined as
\begin{equation}
s(x)=\frac{\exp(z_e)}{\exp(z_c)+\exp(z_e)+\exp(z_n)}.
\label{eq:score}
\end{equation}
Because the score is continuous, a threshold can be set for any target recall on the validation set, which the coarse probabilities of prompted judges do not allow (Section~\ref{sec:thresholds}). We report results for the raw score at threshold 0.5 and use a separate threshold for high-recall screening. To obtain it, we fit a positive sigmoid to the validation log-odds and select the highest threshold achieving 99\% validation recall. We freeze the checkpoint, calibration and threshold before scoring any test skill. At test time, skills scoring at or above the frozen threshold are routed to review.
\section{Experimental Setup}
\label{sec:experiments}

\begin{table*}[t]
\centering
\footnotesize
\setlength{\tabcolsep}{3pt}
\renewcommand{\arraystretch}{1.15}
\newcommand{\mainstat}[2]{\makebox[2.1em][r]{#1}\makebox[1.9em][l]{\if\relax\detokenize{#2}\relax\else\ensuremath{_{\pm#2}}\fi}}
\begin{tabular*}{\textwidth}{@{\extracolsep{\fill}}lrrr>{\columncolor{black!4}}r@{\hspace{12pt}}rrr>{\columncolor{black!4}[\tabcolsep][0pt]}r@{}}
\toprule
& \multicolumn{4}{c}{\textbf{MSD}\quad (Unseen attack structures)} & \multicolumn{4}{c}{\textbf{SD}\quad (Unseen sources)} \\
\cmidrule(lr){2-5}\cmidrule(l){6-9}
& \multicolumn{2}{c}{Default threshold} & & \textbf{99\% recall} & \multicolumn{2}{c}{Default threshold} & & \textbf{99\% recall} \\
Method & Recall $\uparrow$ & FPR $\downarrow$ & AUROC $\uparrow$ & FPR $\downarrow$ & Recall $\uparrow$ & FPR $\downarrow$ & AUROC $\uparrow$ & FPR $\downarrow$ \\
\midrule
\multicolumn{9}{l}{\emph{Off-the-shelf detectors}} \\
Untuned OpenJev-2B & \mainstat{4.2}{} & \mainstat{0.0}{} & 0.678 & 95.5 & \mainstat{0.6}{} & \mainstat{0.0}{} & 0.568 & 99.8 \\
TypeSafe Jev 1.13 & \mainstat{33.8}{} & \mainstat{0.2}{} & 0.841 & 100.0 & \mainstat{11.3}{} & \mainstat{0.0}{} & 0.652 & 100.0 \\
SkillSpector & \mainstat{10.5}{} & \mainstat{0.7}{} & 0.618 & 100.0 & \mainstat{1.3}{} & \mainstat{1.8}{} & 0.433 & 100.0 \\
\midrule
\multicolumn{9}{l}{\emph{LLM judges}} \\
GPT-5.5 (zero-shot) & \mainstat{72.1}{} & \mainstat{0.7}{} & 0.950 & 91.1 & \mainstat{42.8}{} & \mainstat{0.9}{} & 0.914 & 100.0 \\
\quad with 8 examples & \mainstat{87.9}{} & \mainstat{2.5}{} & 0.958 & 95.7 & \mainstat{60.4}{} & \mainstat{2.2}{} & \textbf{0.941} & 79.8 \\
\quad with reasoning & \mainstat{71.1}{} & \mainstat{1.3}{} & 0.953 & 86.8 & \mainstat{33.7}{} & \mainstat{1.1}{} & \underline{0.932} & 100.0 \\
Grok 4.3 & \mainstat{68.1}{} & \mainstat{28.0}{} & 0.707 & 96.6 & \mainstat{36.4}{} & \mainstat{21.3}{} & 0.614 & 97.6 \\
Claude Haiku 4.5 & \mainstat{65.9}{} & \mainstat{2.9}{} & 0.862 & 92.6 & \mainstat{28.8}{} & \mainstat{0.6}{} & 0.732 & 99.8 \\
GPT-5.4 mini & \mainstat{56.3}{} & \mainstat{2.5}{} & 0.862 & 100.0 & \mainstat{23.0}{} & \mainstat{0.9}{} & 0.768 & 100.0 \\
\midrule
\textbf{SkillScreen} (ours) & \mainstat{95.6}{0.6} & \mainstat{3.3}{1.3} & \textbf{0.992} & \textbf{22.6} & \mainstat{97.3}{1.1} & \mainstat{36.0}{4.3} & 0.916 & \textbf{50.5} \\
ModernBERT-large & \mainstat{96.0}{0.4} & \mainstat{4.1}{1.8} & \underline{0.990} & \underline{30.4} & \mainstat{94.1}{4.8} & \mainstat{54.8}{3.8} & 0.769 & \underline{74.7} \\
Word-SVM & \mainstat{94.6}{} & \mainstat{8.9}{} & 0.977 & 45.6 & \mainstat{95.6}{} & \mainstat{62.4}{} & 0.809 & 82.2 \\
\bottomrule
\end{tabular*}
\caption{\textbf{Malicious skill detection on MSD and SD.} Recall and FPR are percentages. Shaded columns give FPR at 99\% test recall. SkillScreen and ModernBERT report five-seed means with subscript standard deviations, using converged SD encoder runs. Bold and underlining mark the best and second-best AUROC and high-recall FPR.}
\label{tab:main}
\end{table*}

% \subsection{Experimental setup}
\paragraph{Split protocols.}
MSD, the main split protocol, contains 6,814 training, 974 validation and 1,947 test skills (1,500 malicious and 447 benign). Its 2,458 training and 1,364 test families are disjoint, and no skill text appears in more than one partition, verbatim or after whitespace normalization.

SD evaluates transfer to unseen sources. It contains 7,510 training, 833 validation and 1,384 test skills (839 malicious and 545 benign), and its test set is drawn from three sources absent from training. Thirty malicious families, covering 38\% of the malicious test skills, also occur in training.

SBR contains 6,812 training, 973 validation and 1,950 test skills. Every source appears in both training and test, and half of the malicious test skills belong to families seen in training. Appendix~\ref{app:composition} gives the source counts, and Appendix~\ref{app:sbr} reports the SBR results.

\paragraph{Baselines.}
ModernBERT-large \citep{warner2024modernbert}, with 395.8M parameters and a newly initialized two-way classification head, serves as a supervised encoder baseline. We fine-tune all its parameters on the raw Markdown at a learning rate of $2\times10^{-5}$. As with SkillScreen, we train for one epoch with five seeds per split, using the same partitions and checkpoint selection rule; on SD a sixth seed replaces one that collapsed. Both models cap inputs at 8,192 tokens, each with its own tokenizer. Appendix~\ref{app:encoder} reports the encoder settings, per-run results and a grid of learning rates on MSD.

Word-SVM follows the lexical baseline of the benchmark, representing the full Markdown of each skill as a 120K-dimensional TF-IDF vector over word unigrams and bigrams with sublinear term frequency. A class-balanced linear SVM classifies a skill as malicious when its margin $\geq$ 0.

Untuned OpenJev-2B uses the same hypothesis without task-specific fine-tuning. We query TypeSafe Jev 1.13 \citep{typesafejev2026}, a commercial NLI API, with the same hypothesis and visible text prefix. SkillSpector \citep{skillspector2026}, a static scanner, is pinned to version 2.12.0 and run with its default risk threshold of 50.

\paragraph{LLM judges.}
Four chat models judge the same visible prefix of each skill with a shared prompt, which defines malicious behaviour and distinguishes it from authorized developer and security workflows. Each request marks the skill as untrusted and asks for a boolean verdict and a probability in JSON. We collect one answer per record and classify a skill as malicious when the reported probability is at least 0.5.

GPT-5.5 and GPT-5.4 mini run without reasoning tokens, and Claude Haiku 4.5 without extended thinking. Grok 4.3 is accessed through a reseller; although our requests set reasoning effort to none, its responses still reported reasoning tokens. We also evaluate two GPT-5.5 conditions on both protocols, one with medium reasoning and one with eight labelled examples in the prompt, drawn from the training partition of the protocol under test. All prompts and examples are fixed before the corresponding test requests are sent. Appendix~\ref{app:prompt} details the request settings.

\paragraph{Metrics and uncertainty.}
Default decisions use probability $\ge 0.5$, SVM margin $\ge 0$ and scanner risk $>50$.
We report recall on malicious skills, the false-positive rate (FPR) on benign skills, the area under the receiver operating characteristic curve (AUROC) and average precision (AP). SkillScreen and ModernBERT-large are trained with five seeds per split (six for ModernBERT-large on SD), the plain-base and unrelated-hypothesis ablations with five seeds and the 0.8B ablation with two. Each 10\%, 25\% and 50\% data condition comprises nine runs (three subset draws, each with three seeds), and all other trained variants use a single seed.

Tables report means and standard deviations across seeds. We compute 95\% paired bootstrap intervals from 2,000 resamples of the test records \citep{berg-kirkpatrick-etal-2012-empirical, dror-etal-2018-hitchhikers}. Each resample fixes the class sizes and uses the same record indices for all methods. For multi-seed methods, we average the estimates across seeds within each resample. These intervals capture variation across records from the observed sources, while the seed spread is reported separately.

To compare methods at a shared recall target, we read each method's threshold off a sweep over its own test scores and recompute it within every bootstrap replicate, so the interval for the FPR at 99\% recall reflects the few lowest-scoring malicious skills that set the threshold (Appendix~\ref{app:review}). The high-recall thresholds used for decisions, in contrast, are selected on the validation set.

\section{Results}
\subsection{Main comparison on MSD}
On unseen attack families, SkillScreen combines high recall with low false-positive rates (Table~\ref{tab:main}). The shaded columns compare the false-positive cost of reaching 99\% recall, the operating point targeted by our screening task. At the default threshold, SkillScreen lowers the FPR by 5.6 points relative to Word-SVM on the same resampled test records ($[-8.5,-3.0]$) at comparable recall ($+1.0$, $[-0.1,+2.1]$) and raises AUROC by 0.015 ($[+0.010,+0.021]$). SkillScreen generalizes across attack structures: none of the 1,364 MSD test families occurs in training, 90\% of them are singletons, and on the random split recall is nearly the same for malicious skills from seen (98.8\%) and unseen (97.4\%) families (five seeds; Appendix~\ref{app:review}).

Task training is what turns the NLI model into a detector. At threshold 0.5, untuned OpenJev-2B detects 4\% of malicious skills (AUROC 0.678) and the commercial NLI API 34\% (AUROC 0.841), versus 95.1--96.5\% across the five SkillScreen seeds.

ModernBERT-large is the strongest supervised baseline. Near the default threshold the two models are on par: matching recall at 95.6\% gives equal FPR (paired difference $0.0$ points, $[-1.8,+2.0]$). They separate at the recall that screening requires: at 99\% recall, SkillScreen needs 22.6\% FPR, against 30.4\% for ModernBERT and 45.6\% for Word-SVM. Section~\ref{sec:thresholds} examines this high-recall advantage and its uncertainty, and Section~\ref{sec:sd} shows that it widens under source shift.

\subsection{High-recall operating points}
\label{sec:thresholds}
High-recall operating points separate models with similar AUROC (Table~\ref{tab:main}, shaded MSD column; Figure~\ref{fig:curves}(b) in Appendix~\ref{app:sdfig}). At 99\% recall, FPR ranges from 19.9--24.6\% across SkillScreen seeds and 25.1--36.2\% across ModernBERT seeds. Every SkillScreen seed requires a lower FPR than every encoder run. Bootstrap intervals for the seed means are $[13.5,37.0]$ and $[21.2,51.4]$, respectively, and $[30.0,66.2]$ for Word-SVM. Relative to SkillScreen, the paired FPR differences are $+7.8$ points for the encoder ($[-3.4, +22.1]$) and $+23.0$ points for Word-SVM ($[+6.8, +39.8]$). The ordering holds for every pair of seeds; the record-level interval still includes zero on MSD, and the gap becomes clear under source shift (Section~\ref{sec:sd}). At 95.6\% recall, SkillScreen requires an FPR of 2.0--4.0\%, compared with 13.6\% for Word-SVM and 34.7\% for GPT-5.5. The prompted judges have no usable 99\%-recall operating point: their probabilities take only 13--60 distinct values, so the first threshold that reaches 99\% recall flags about 87--100\% of benign skills (Table~\ref{tab:main}). Screening requires a score that can be thresholded finely, which a trained scorer provides by construction.

Thresholds frozen on the validation set preserve this operating point on the test set. At thresholds chosen for 99\% validation recall, SkillScreen attains 98.7--99.1\% test recall at 20.1--23.9\% FPR. The ModernBERT-large runs, which start from a higher validation FPR at 99\% recall (27.7\% on average, compared with 22.8\% for SkillScreen), attain 98.5--99.1\% test recall at 25.1--38.0\% FPR, and Word-SVM attains 99.3\% test recall at 60.6\% FPR (Appendices~\ref{app:gates} and~\ref{app:encoder}).

On SBR, SkillScreen reaches 99\% test recall at 6.6\% mean FPR, compared with 12.1\% for ModernBERT-large (Appendix~\ref{app:sbr}).

\paragraph{Prompted judges.}
Neither in-context examples nor reasoning closes the gap to SkillScreen (Table~\ref{tab:main}). Eight labelled examples raise GPT-5.5's recall from 72.1\% to 87.9\%, yet at SkillScreen's mean recall it still requires an FPR of 28.2\%, and medium reasoning leaves its recall at 71.1\%.

SkillScreen also runs locally at no API cost: on a single 10\,GB MIG slice, it scores 1,000 skills in about 4.5 minutes, whereas the judges cost USD\,4.3--8.8 per 1,000 skills for the GPT-5.5 and Grok conditions and USD\,0.8--1.2 for the smaller models. Appendix~\ref{app:prompt} details the settings and source-level results of the judges.

\begin{figure*}[t]
\centering
\includegraphics[width=\textwidth]{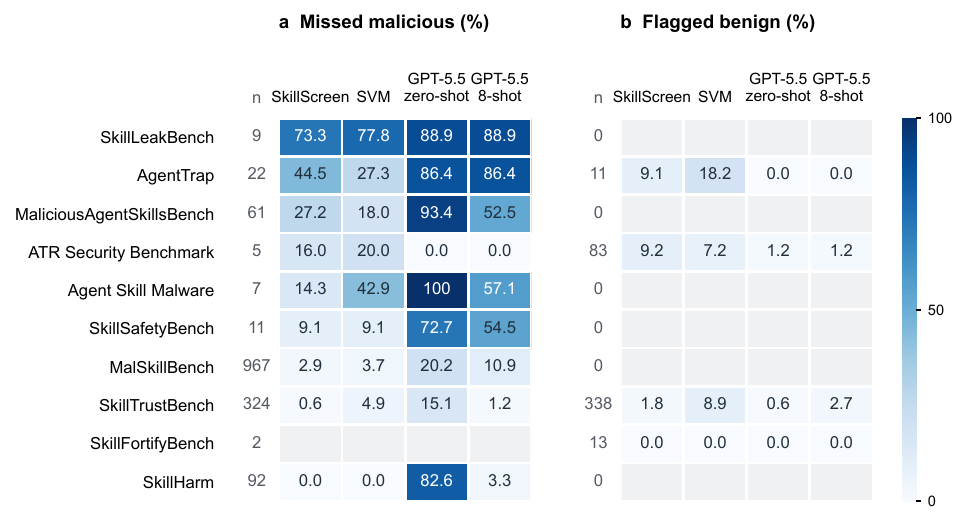}
\caption{Source-level error patterns underlying the MSD results. At the default thresholds, panel (a) shows the malicious miss rate and panel (b) the benign FPR; darker cells indicate more errors. The adjacent $n$ gives the number of records in the corresponding class. SkillScreen values are averaged over five seeds. Blank grey cells mark sources with fewer than five records in that class. Sources are ordered by the miss rate of SkillScreen.}
\label{fig:sources}
\end{figure*}

\paragraph{Error analysis.}
On MSD, SkillScreen identifies most of the malicious skills that the strongest judge misses: every seed detects 346 of the 419 malicious skills missed by GPT-5.5, whereas GPT-5.5 detects nine of the 42 malicious skills missed by all five seeds. Errors shared by all seeds are few (42 missed malicious skills and four flagged benign skills), and several of the shared misses contain no malicious evidence in their Markdown at all. Their labels may instead rest on package code that is absent from \texttt{SKILL.md} (Figure~\ref{fig:sources}; Appendix~\ref{app:errors}).

\subsection{Benign source shift}
\label{sec:sd}

On unseen sources (Table~\ref{tab:main}, SD), SkillScreen keeps the highest recall, 97.3\% at 0.5 and 98.9--99.6\% at the frozen gates, and the lowest FPR at 99\% recall: 50.5\% (46--56\% across seeds), against 74.7\% for the converged ModernBERT-large runs ($+24.1$ points, $[+14.4, +32.7]$), 79.8\% for the best prompted judge and 82.2\% for Word-SVM (Figure~\ref{fig:sdcurves}(b) in Appendix~\ref{app:sdfig}). GPT-5.5 ranks skills comparably (AUROC 0.914 zero-shot and 0.941 with eight SD training examples, versus 0.916) but lacks a usable high-recall operating point: at 95.6\% recall, the zero-shot judge flags 78.5\% of benign skills and the eight-example judge 65.7\%, compared with 24--33\% for SkillScreen (Appendix~\ref{app:sdfig}).

False positives concentrate in the two benign test sources, both unlike any training source. SkillFortifyBench is generated from a synthetic template whose benign and malicious variants differ only in their commands (\texttt{df -h} versus \texttt{rm -rf}). Every SkillScreen seed and converged encoder run flags all 90 benign and 90 malicious variants alike, whereas GPT-5.5 reads the commands and flags none of the benign variants. Of the 455 benign skills from the ATR Security Benchmark, which describe security tooling, SkillScreen flags 17--30\% at 0.5, compared with 46\% for the encoder and 1.1\% for GPT-5.5 (Appendix~\ref{app:errors}).

The shift lies in the benign distribution. SkillScreen's thresholds are chosen at 8--11\% validation FPR on benign skills from the training sources, 165 of 169 of them from SkillTrustBench, which is absent from the test set; on the unseen test sources, they flag 56--64\% of benign skills, compared with 62--84\% for the encoder's gates and 92.3\% for that of Word-SVM. On MSD, where benign sources are shared, validation and test FPR agree (20--28\% and 20--24\%), as does calibration (Appendix~\ref{app:calibration}). These results point to benign training data from more sources as the remedy; a precise judge cannot filter SkillScreen's flags without inheriting the judge's recall (42.7\% on SD; Appendix~\ref{app:review}).

\subsection{Training choices}
\paragraph{Scanner context.}
\label{sec:context}
The skill text alone suffices. Two variants append the SkillSpector output to the same skill-text prefix: the full-context variant adds the scanner score, verdict, severity, categories, findings and completeness fields, and the findings-only variant omits the final score and verdict. They reach 95.1\% recall at 2.5\% FPR and 95.4\% at 1.3\%, respectively, with paired intervals relative to SkillScreen seed 42 that include zero (Appendix~\ref{app:context}), while scanning adds 817\,s of runtime on the MSD test set; on SBR, both variants trail SkillScreen at their high-recall thresholds (Appendix~\ref{app:sbr}).

\begin{table}[t]
\centering
\footnotesize
\setlength{\tabcolsep}{3pt}
\resizebox{\columnwidth}{!}{%
\begin{tabular}{lrrrr}
\toprule
Variant & Recall & FPR & AUROC & FPR@99 \\
\midrule
SkillScreen (5 seeds) & 95.6$\pm$0.6 & 3.3$\pm$1.3 & 0.992 & 22.6 \\
Plain base (5 seeds) & 96.9$\pm$0.8 & 4.1$\pm$1.6 & 0.992 & 21.1 \\
Paraphrased hypothesis & 95.1 & 1.3 & 0.993 & 21.0 \\
Unrelated hypothesis (5 seeds) & 95.7$\pm$0.8 & 3.3$\pm$1.2 & 0.990 & 28.9 \\
Empty hypothesis & 95.1 & 2.0 & 0.988 & 44.7 \\
0.8B base (2 seeds) & 95.4$\pm$0.0 & 2.9$\pm$0.9 & 0.991 & 26.0 \\
\bottomrule
\end{tabular}}
\caption{Ablations on MSD. Recall and FPR are measured at probability 0.5; the last column reports the benign FPR at 99\% test recall from the threshold sweep of each run. All rates are seed means in percent. The plain base is Qwen3.5-2B with a randomly initialized head; the paraphrased and empty hypotheses use a single seed. Data, epoch budget and adapter settings are held fixed. Appendix~\ref{app:review} reports paired intervals relative to SkillScreen; at 99\% recall, only the empty hypothesis is clearly worse ($+22.1$ points, $[+2.0, +40.4]$).}
\label{tab:ablations}
\end{table}

\paragraph{Initialization, hypothesis and size.}
\label{sec:recipe}
Task training is robust to the starting point (Table~\ref{tab:ablations}). Trained from plain Qwen3.5-2B with a randomly initialized three-way head, whose untrained test AP of 0.76 at a malicious prevalence of 77\% is close to chance, the scorer reaches 96.9\,$\pm$\,0.8\% recall at 4.1\,$\pm$\,1.6\% FPR over five seeds (AUROC 0.992, AP 0.998), against 95.6\,$\pm$\,0.6\% at 3.3\,$\pm$\,1.3\% for SkillScreen. Five seeds with the unrelated hypothesis match SkillScreen (95.7\,$\pm$\,0.8\% recall at 3.3\,$\pm$\,1.2\% FPR), and the single paraphrased and empty-hypothesis runs reach 95.1\% recall at 1.3\% and 2.0\% FPR, with paired intervals relative to SkillScreen seed 42 that include zero.

The content of the hypothesis matters where the decision margin is narrow. At 99\% recall, an empty hypothesis adds 22.1 FPR points ($[+2.0, +40.4]$), whereas the paraphrase ($-1.6$, $[-11.9, +4.7]$) and the plain base ($-1.5$, $[-11.9, +8.0]$) stay within noise of SkillScreen and the unrelated hypothesis lies in between ($+6.3$, $[-3.6, +19.3]$). Before training, a hypothesis that names the task already ranks skills better: validation AP is 0.903 for the original and 0.892 for the paraphrase, against 0.784 for the unrelated and 0.716 for the empty hypothesis.

Model size matters at high recall: OpenJev-0.8B matches the 2B model at the fixed threshold (95.4\% recall at 2.9\% FPR) but needs 30--34\% FPR at the high-recall thresholds, against 20--24\%, while running only 10--20\% faster.

\begin{figure*}[t]
\centering
\includegraphics[width=\textwidth]{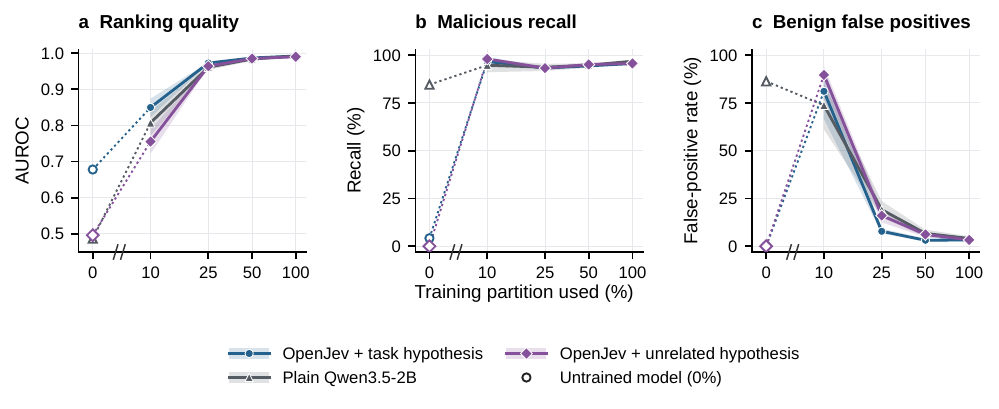}
\caption{With limited labels, the full recipe learns most: OpenJev with the task-naming hypothesis leads in AUROC at 10\% of the labels and flags about half as many benign skills as either alternative at 25--50\%. Panels show (a) AUROC, (b) recall and (c) FPR on MSD, with recall and FPR measured at probability 0.5. Each point at 10\%, 25\% and 50\% averages nine runs (three stratified subset draws with three seeds each), and each full-data point averages five seeds. Shaded bands are 95\% intervals of the mean from resampling the runs; hollow markers at 0\% show each starting model before task training. All runs are trained for one epoch.}
\label{fig:dataeff}
\end{figure*}
\subsection{Learning from fewer skills}
\label{sec:dataeff}
Training with fewer labels shows what each part of the recipe contributes (Figure~\ref{fig:dataeff}). We use stratified 10\%, 25\% and 50\% subsets of the training set (681, 1,704 and 3,407 skills, respectively), drawing three subsets at each fraction and training three seeds on each. At 10\%, the task-naming hypothesis lifts ranking quality: the original configuration reaches an AUROC of 0.850\,$\pm$\,0.043, versus 0.755\,$\pm$\,0.055 for the unrelated hypothesis, and outperforms it in all nine runs paired by subset and seed (mean AUROC difference $+0.094$, two-sided sign test $p=0.004$); its edge over the plain base, at 0.807\,$\pm$\,0.058, is smaller ($+0.030$; five of eight pairs, $p=0.73$). From 25\% of the labels on, the full recipe flags about half as many benign skills as either alternative at a similar recall of 93--95\%: 7.8\% at 25\% and 3.1\% at 50\%, against 16.1--19.2\% and 6.2--6.7\%, with non-overlapping 95\% intervals. Its ranking edge narrows but stays consistent: at 25\%, it outperforms either alternative in all eight seed-matched pairs (mean AUROC difference $+0.010$ against the unrelated hypothesis and $+0.013$ against the plain base; sign test $p=0.008$ each; Appendix~\ref{app:lowdata}). Both the NLI initialization and the task-naming hypothesis thus improve learning from limited labels.
\section{Related work}

\paragraph{Attacks on agents.}
Indirect prompt injection embeds adversarial instructions in content consumed by LLM applications \citep{greshake2023notwhat, perez2022ignorepreviouspromptattack, liu2025promptinjectionattackllmintegrated}. InjecAgent, AgentDojo and Agent Security Bench test these attacks and defences through agent interactions \citep{zhan2024injecagentbenchmarkingindirectprompt, debenedetti2024agentdojo, zhang2025agentsecuritybenchasb}. MCP tool descriptions introduce similar risks \citep{radosevich2025mcpsafetyauditllms}. We screen skills before agents use them.

\paragraph{Malicious skills and detection.}
\citet{liu2026domentionuserdetecting} document credential theft and agent manipulation in public skill registries. SkillInject evaluates agents against attacks embedded in skill files \citep{schmotz2026skillinjectmeasuringagentvulnerability}. MalSkillBench verifies malicious behaviour at runtime and evaluates detectors across code and instruction attacks \citep{guo2026malskillbenchruntimeverifiedbenchmarkmalicious}. MaliciousSkillBench consolidates public sources and supplies structural and source-disjoint protocols \citep{wang2026maliciousskillbench}. Under these protocols, we show that a small trained scorer screens skills at high recall and trace where source shift moves its false positives.

\paragraph{Malicious packages.}
Software supply-chain research studies attacks distributed through third-party packages \citep{ohm2020backstabbersknifecollectionreview}. Classifiers can detect malicious packages from metadata and code \citep{sejfia2022practicalautomateddetectionmalicious}. Agent skills add instructions that an LLM interprets directly. SkillSpector applies static rules to skill packages \citep{skillspector2026}. SkillScreen instead learns to screen these instructions from labelled skills, detecting 95.6\% of malicious skills where static rules catch 10.5\% (Table~\ref{tab:main}).

\paragraph{Decision models.}
Decision models return probabilities over predefined outcomes \citep{typesafejev2026,ibrahim2026decisionmodels}. Llama Guard and ShieldGemma specialize language models for safety screening \citep{inan2023llamaguard,zeng2024shieldgemma}.

\paragraph{NLI and model-based judgments.}
NLI supports zero-shot and few-shot classification by expressing labels as hypotheses \citep{yin2019benchmarking, wang2021entailmentfewshotlearner}. SkillScreen adapts this interface to skill detection through LoRA \citep{hu2021lora}. Work on LLM judges studies prompted assessments and their biases \citep{zheng2023judging, wang2023largelanguagemodelsfair}. LLM monitors apply such judgments to oversight, flagging reward hacking, sabotage and other misbehaviour in model outputs and agent trajectories \citep{greenblatt2024aicontrolimprovingsafety, baker2025monitoringreasoningmodelsmisbehavior, kutasov2025shadearenaevaluatingsabotagemonitoring, yang2026automonitorbenchevaluatingreliabilityllmbased, quan2026rhdetectunifiedbenchmarkreward}. On the same visible skill text, we show that task-trained scoring reaches high attack recall at a fraction of the false-positive cost of prompted judgments.
\section{Conclusion}
SkillScreen shows that high-recall screening of agent skills does not need a frontier model. A local 2B decision model, trained with one epoch of LoRA on 6.8K labelled skills, flags 20--25\% of benign skills at 99\% recall on attack families unseen in training, where a word SVM needs 46\% and prompted LLM judges flag about 87\% or more, and thresholds frozen on validation data hold this operating point on test. Among trained detectors, its advantage is largest where screening is hardest: on unseen skill sources, its lead over a ModernBERT-large encoder fully fine-tuned for the same single epoch widens to 24 points at 99\% recall. Task training drives the gain; a hypothesis that names the task sharpens it at high recall, and with 25--50\% of the labels the full recipe flags half as many benign skills as either ablation. The Markdown model supports initial registry screening on a single GPU slice. The same decision-model formulation extends to supporting files within skill packages (Appendix~\ref{app:sbr}). We release both models with data and inference code (Appendix~\ref{app:repro}).

\section*{Limitations}
Our main benchmark comparisons use \texttt{SKILL.md}. We evaluate the Markdown model against an attacker who publishes a skill, using fixed rewrites rather than an adaptive attack. Paraphrasing 200 malicious test skills leaves its recall essentially unchanged. Moving their commands into a referenced script and neutralizing the prose takes the evidence out of \texttt{SKILL.md}: recall at the frozen gate falls from 99.8\% to 87.0\%, and at 0.5 from 96.6\% to 67.5\% (Appendix~\ref{app:robustness}). Every detector that reads only \texttt{SKILL.md} shares this limit. The package variant in Appendix~\ref{app:sbr} extends the input to supporting files. We applied the rewrite to malicious skills and to SkillScreen only.

We keep the official benchmark labels, some of which depend on package code or behaviour absent from \texttt{SKILL.md} and thus cap what any text-only detector can recover. MSD holds out malicious families while its benign skills share sources across partitions, and SD contains two benign test sources. Recall varies by source (macro-average 79\% against 95.6\% overall), with wide intervals on the small sources.

On MSD we report the encoder's GPU rerun, the stronger of its two five-seed sets; Appendix~\ref{app:encoder} gives both, including the two collapsed runs. Judge comparisons use one prompt, one demonstration set, the stated provider settings and the probabilities the judges report; log-probabilities or repeated sampling would give the judges finer-grained scores at 99\% recall. The one-epoch schedule varies the number of updates with data size, the scanner-context variants and the paraphrased and empty hypotheses use one seed each, and the size ablation also changes the NLI checkpoint version.

At 99\% recall, SkillScreen is a triage stage rather than an alarm: at an FPR of about 21\%, precision is 34\% when 10\% of a stream is malicious and 4.5\% when 1\% is. Deployment estimates also depend on the source mix and on the model's pretraining overlap with public skills.

\section*{Ethics statement}
The inputs are public skill texts. We used no private corpus or credentials and executed no skill-supplied commands. The judge prompt marks all skill text as untrusted. SkillScreen routes suspicious skills to human review. False positives add review work, and missed malicious skills can expose users to attacks. Releasing the scorer and its per-skill predictions could also help attackers probe for evasions; the most effective rewrite we tested (Appendix~\ref{app:robustness}) needs no model access, and we judge that the defensive value outweighs this risk.

Third-party skill texts retain their upstream licences, and we do not redistribute them: our scripts retrieve them from the pinned MaliciousSkillBench release, whose metadata and split manifests are licensed under CC BY 4.0. Released prediction files contain benchmark identifiers, labels and scores, with source names and family identifiers in a separate metadata file; raw skill text and API responses are excluded. The adapters we release build on OpenJev-2B (MIT licence) and Qwen3.5-2B (Apache 2.0), whose licences permit this release.

\section*{Acknowledgments}
We thank the Nebula team at TrendAI for discussions about the security of agent skills during its industry project with the University of Toronto’s MScAC program. We also thank Haoxuan Huang and Shiyi Yang of the University of Toronto for their work with Nebula on that project.

\bibliography{skilljev}

\begin{thebibliography}{45}
\providecommand{\natexlab}[1]{#1}

\bibitem[{{Agent Skills project}(2026)}]{agentskills2026spec}
{Agent Skills project}. 2026.
\newblock \href {https://agentskills.io/specification} {{Agent Skills}
  specification}.
\newblock Online specification, checked 2026-10-07.

\bibitem[{{Agent-Threat-Rule}(2026)}]{atrskillbenchmark2026}
{Agent-Threat-Rule}. 2026.
\newblock \href
  {https://huggingface.co/datasets/Agent-Threat-Rule/atr-skill-benchmark} {{ATR
  Skill-Security Benchmark}}.
\newblock Hugging Face dataset.

\bibitem[{{AlexWortega}(2026)}]{openjev2026}
{AlexWortega}. 2026.
\newblock \href {https://huggingface.co/AlexWortega/openjev} {{OpenJev}: Open
  natural-language-inference checkpoints (qwen3.5-2b-nli-v5)}.
\newblock Hugging Face model repository, revision
  a20448012c213128955ca0c693e7c943865cab77.

\bibitem[{Baker et~al.(2025)Baker, Huizinga, Gao, Dou, Guan, Madry, Zaremba,
  Pachocki, and Farhi}]{baker2025monitoringreasoningmodelsmisbehavior}
Bowen Baker, Joost Huizinga, Leo Gao, Zehao Dou, Melody~Y. Guan, Aleksander
  Madry, Wojciech Zaremba, Jakub Pachocki, and David Farhi. 2025.
\newblock \href {https://arxiv.org/abs/2503.11926} {Monitoring reasoning models
  for misbehavior and the risks of promoting obfuscation}.
\newblock \emph{Preprint}, arXiv:2503.11926.

\bibitem[{Berg-Kirkpatrick et~al.(2012)Berg-Kirkpatrick, Burkett, and
  Klein}]{berg-kirkpatrick-etal-2012-empirical}
Taylor Berg-Kirkpatrick, David Burkett, and Dan Klein. 2012.
\newblock \href {https://aclanthology.org/D12-1091/} {An empirical
  investigation of statistical significance in {NLP}}.
\newblock In \emph{Proceedings of the 2012 Joint Conference on Empirical
  Methods in Natural Language Processing and Computational Natural Language
  Learning}, pages 995--1005, Jeju Island, Korea. Association for Computational
  Linguistics.

\bibitem[{Bhardwaj(2026)}]{bhardwaj2026skillfortify}
Varun~Pratap Bhardwaj. 2026.
\newblock \href {https://arxiv.org/abs/2603.00195} {Formal analysis and supply
  chain security for agentic {AI} skills}.
\newblock \emph{Preprint}, arXiv:2603.00195.

\bibitem[{Chen et~al.(2026)Chen, Zhang, Liu, Deng, Li, Zhang, Ning, Zhang, Ma,
  and Li}]{chen2026skillleakbench}
Zhihao Chen, Ying Zhang, Yi~Liu, Gelei Deng, Yuekang Li, Yanjun Zhang, Jianting
  Ning, Leo~Yu Zhang, Lei Ma, and Zhiqiang Li. 2026.
\newblock \href {https://arxiv.org/abs/2604.03070} {How your credentials are
  leaked by {LLM} agent skills: An empirical study}.
\newblock \emph{Preprint}, arXiv:2604.03070.

\bibitem[{{cuhk-zhuque}(2026)}]{skilltrustbench2026}
{cuhk-zhuque}. 2026.
\newblock \href {https://huggingface.co/datasets/cuhk-zhuque/SkillTrustBench}
  {{SkillTrustBench}}.
\newblock Hugging Face dataset, version v1.0.

\bibitem[{Debenedetti et~al.(2024)Debenedetti, Zhang, Balunović,
  Beurer-Kellner, Fischer, and Tramèr}]{debenedetti2024agentdojo}
Edoardo Debenedetti, Jie Zhang, Mislav Balunović, Luca Beurer-Kellner, Marc
  Fischer, and Florian Tramèr. 2024.
\newblock \href {https://doi.org/10.52202/079017-2636} {{AgentDojo}: A dynamic
  environment to evaluate prompt injection attacks and defenses for {LLM}
  agents}.
\newblock In \emph{Advances in Neural Information Processing Systems 37}, pages
  82895--82920.

\bibitem[{Dror et~al.(2018)Dror, Baumer, Shlomov, and
  Reichart}]{dror-etal-2018-hitchhikers}
Rotem Dror, Gili Baumer, Segev Shlomov, and Roi Reichart. 2018.
\newblock \href {https://doi.org/10.18653/v1/P18-1128} {The hitchhiker{'}s
  guide to testing statistical significance in natural language processing}.
\newblock In \emph{Proceedings of the 56th Annual Meeting of the Association
  for Computational Linguistics (Volume 1: Long Papers)}, pages 1383--1392,
  Melbourne, Australia. Association for Computational Linguistics.

\bibitem[{Feng et~al.(2026)Feng, Ding, Tan, Zheng, Guo, Li, Zhai, Li, and
  Huang}]{feng2026skilltrojan}
Yunhao Feng, Yifan Ding, Yingshui Tan, Boren Zheng, Yanming Guo, Xiaolong Li,
  Kun Zhai, Yishan Li, and Wenke Huang. 2026.
\newblock \href {https://arxiv.org/abs/2604.06811} {{SkillTrojan}: Backdoor
  attacks on skill-based agent systems}.
\newblock \emph{Preprint}, arXiv:2604.06811.

\bibitem[{Greenblatt et~al.(2024)Greenblatt, Shlegeris, Sachan, and
  Roger}]{greenblatt2024aicontrolimprovingsafety}
Ryan Greenblatt, Buck Shlegeris, Kshitij Sachan, and Fabien Roger. 2024.
\newblock \href {https://proceedings.mlr.press/v235/greenblatt24a.html} {{AI}
  control: Improving safety despite intentional subversion}.
\newblock In \emph{Proceedings of the 41st International Conference on Machine
  Learning}, volume 235 of \emph{Proceedings of Machine Learning Research},
  pages 16295--16336. PMLR.

\bibitem[{Greshake et~al.(2023)Greshake, Abdelnabi, Mishra, Endres, Holz, and
  Fritz}]{greshake2023notwhat}
Kai Greshake, Sahar Abdelnabi, Shailesh Mishra, Christoph Endres, Thorsten
  Holz, and Mario Fritz. 2023.
\newblock \href {https://doi.org/10.1145/3605764.3623985} {Not what you've
  signed up for: Compromising real-world {LLM}-integrated applications with
  indirect prompt injection}.
\newblock In \emph{Proceedings of the 16th ACM Workshop on Artificial
  Intelligence and Security}, pages 79--90. ACM.

\bibitem[{Guo et~al.(2017)Guo, Pleiss, Sun, and
  Weinberger}]{guo2017calibrationmodernneuralnetworks}
Chuan Guo, Geoff Pleiss, Yu~Sun, and Kilian~Q. Weinberger. 2017.
\newblock \href {https://proceedings.mlr.press/v70/guo17a.html} {On calibration
  of modern neural networks}.
\newblock In \emph{Proceedings of the 34th International Conference on Machine
  Learning}, volume~70 of \emph{Proceedings of Machine Learning Research},
  pages 1321--1330. PMLR.

\bibitem[{Guo et~al.(2026)Guo, Zeng, Liu, Jia, Xu, Tang, Fang, and
  Liu}]{guo2026malskillbenchruntimeverifiedbenchmarkmalicious}
Wenbo Guo, Wei Zeng, Chengwei Liu, Xiaojun Jia, Yijia Xu, Lei Tang, Yong Fang,
  and Yang Liu. 2026.
\newblock \href {https://arxiv.org/abs/2606.07131} {{MalSkillBench}: A
  runtime-verified benchmark of malicious agent skills}.
\newblock \emph{Preprint}, arXiv:2606.07131.

\bibitem[{Hu et~al.(2022)Hu, Shen, Wallis, Allen-Zhu, Li, Wang, Wang, and
  Chen}]{hu2021lora}
Edward~J. Hu, Yelong Shen, Phillip Wallis, Zeyuan Allen-Zhu, Yuanzhi Li, Shean
  Wang, Lu~Wang, and Weizhu Chen. 2022.
\newblock \href {https://openreview.net/forum?id=nZeVKeeFYf9} {{LoRA}: Low-rank
  adaptation of large language models}.
\newblock In \emph{International Conference on Learning Representations}.

\bibitem[{Ibrahim and Zaki(2026)}]{ibrahim2026decisionmodels}
Hazem Ibrahim and Yasir Zaki. 2026.
\newblock \href {https://arxiv.org/abs/2609.24574} {Evaluating decision models
  for text annotation in computational social science}.
\newblock \emph{Preprint}, arXiv:2609.24574.

\bibitem[{Inan et~al.(2023)Inan, Upasani, Chi, Rungta, Iyer, Mao, Tontchev, Hu,
  Fuller, Testuggine, and Khabsa}]{inan2023llamaguard}
Hakan Inan, Kartikeya Upasani, Jianfeng Chi, Rashi Rungta, Krithika Iyer,
  Yuning Mao, Michael Tontchev, Qing Hu, Brian Fuller, Davide Testuggine, and
  Madian Khabsa. 2023.
\newblock \href {https://arxiv.org/abs/2312.06674} {{Llama Guard}: {LLM}-based
  input-output safeguard for human-{AI} conversations}.
\newblock \emph{Preprint}, arXiv:2312.06674.

\bibitem[{Jin et~al.(2026)Jin, Wang, Wei, Wang, Zeng, Zhang, Yang, Qu, Hu, and
  Xu}]{jin2026skillsafetybench}
Chang Jin, An~Wang, Zeming Wei, Kai Wang, Biaojie Zeng, Qiaosheng Zhang, Chao
  Yang, Jingjing Qu, Xia Hu, and Xingcheng Xu. 2026.
\newblock \href {https://arxiv.org/abs/2605.12015} {{SkillSafetyBench}:
  Evaluating agent safety under skill-facing attack surfaces}.
\newblock \emph{Preprint}, arXiv:2605.12015.

\bibitem[{Kutasov et~al.(2025)Kutasov, Sun, Colognese, van~der Weij, Petrini,
  Zhang, Hughes, Deng, Sleight, Tracy, Shlegeris, and
  Benton}]{kutasov2025shadearenaevaluatingsabotagemonitoring}
Jonathan Kutasov, Yuqi Sun, Paul Colognese, Teun van~der Weij, Linda Petrini,
  Chen Bo~Calvin Zhang, John Hughes, Xiang Deng, Henry Sleight, Tyler Tracy,
  Buck Shlegeris, and Joe Benton. 2025.
\newblock \href {https://arxiv.org/abs/2506.15740} {{SHADE-Arena}: Evaluating
  sabotage and monitoring in {LLM} agents}.
\newblock \emph{Preprint}, arXiv:2506.15740.

\bibitem[{Liu et~al.(2026)Liu, Chen, Zhang, Deng, Li, Ning, and
  Zhang}]{liu2026domentionuserdetecting}
Yi~Liu, Zhihao Chen, Yanjun Zhang, Gelei Deng, Yuekang Li, Jianting Ning, and
  Leo~Yu Zhang. 2026.
\newblock \href
  {https://www.usenix.org/conference/usenixsecurity26/presentation/liu-yi}
  {``{Do} not mention this to the user'': Detecting and understanding malicious
  agent skills in the wild}.
\newblock In \emph{35th USENIX Security Symposium (USENIX Security 26)}, pages
  1727--1746, Baltimore, MD. USENIX Association.

\bibitem[{Liu et~al.(2025)Liu, Deng, Li, Wang, Wang, Wang, Zhang, Liu, Wang,
  Zheng, Zhang, and Liu}]{liu2025promptinjectionattackllmintegrated}
Yi~Liu, Gelei Deng, Yuekang Li, Kailong Wang, Zihao Wang, Xiaofeng Wang,
  Tianwei Zhang, Yepang Liu, Haoyu Wang, Yan Zheng, Leo~Yu Zhang, and Yang Liu.
  2025.
\newblock \href {https://arxiv.org/abs/2306.05499} {Prompt injection attack
  against {LLM}-integrated applications}.
\newblock \emph{Preprint}, arXiv:2306.05499.

\bibitem[{{Model Context Protocol project}(2025)}]{mcp2025spec}
{Model Context Protocol project}. 2025.
\newblock \href {https://modelcontextprotocol.io/specification/2025-06-18}
  {{Model Context Protocol} specification, revision 2025-06-18}.
\newblock Online specification, checked 2026-10-07.

\bibitem[{Ning et~al.(2026)Ning, Zhang, Lal, Gou, Li, Ruan, Ye, Gupta, Yang,
  Su, and Sun}]{ning2026skillharm}
Yuting Ning, Zhehao Zhang, Yash~Kumar Lal, Boyu Gou, Junyi Li, Weitong Ruan,
  Chentao Ye, Rahul Gupta, Diyi Yang, Yu~Su, and Huan Sun. 2026.
\newblock \href {https://arxiv.org/abs/2606.02540} {{SkillHarm}:
  Lifecycle-aware skill-based attacks via automated construction}.
\newblock \emph{Preprint}, arXiv:2606.02540.

\bibitem[{{NVIDIA}(2026)}]{skillspector2026}
{NVIDIA}. 2026.
\newblock \href {https://github.com/NVIDIA/SkillSpector} {{SkillSpector}:
  Static analysis for {AI} agent skills}.
\newblock GitHub repository, commit 35270064.

\bibitem[{Ohm et~al.(2020)Ohm, Plate, Sykosch, and
  Meier}]{ohm2020backstabbersknifecollectionreview}
Marc Ohm, Henrik Plate, Arnold Sykosch, and Michael Meier. 2020.
\newblock \href {https://doi.org/10.1007/978-3-030-52683-2_2} {Backstabber's
  knife collection: A review of open source software supply chain attacks}.
\newblock In \emph{Detection of Intrusions and Malware, and Vulnerability
  Assessment ({DIMVA} 2020)}, Lecture Notes in Computer Science, pages 23--43.
  Springer.

\bibitem[{Perez and Ribeiro(2022)}]{perez2022ignorepreviouspromptattack}
Fábio Perez and Ian Ribeiro. 2022.
\newblock \href {https://arxiv.org/abs/2211.09527} {Ignore previous prompt:
  Attack techniques for language models}.
\newblock In \emph{NeurIPS ML Safety Workshop}.

\bibitem[{Quan et~al.(2026)Quan, Opryshko, Subramani, and
  Gilitschenski}]{quan2026rhdetectunifiedbenchmarkreward}
Junwei Quan, Evgenii Opryshko, Rohan Subramani, and Igor Gilitschenski. 2026.
\newblock \href {https://arxiv.org/abs/2610.10947} {{RH-Detect}: A unified
  benchmark for reward hacking detection}.
\newblock \emph{Preprint}, arXiv:2610.10947.

\bibitem[{{Qwen Team}(2026)}]{qwen2026qwen35}
{Qwen Team}. 2026.
\newblock \href {https://huggingface.co/Qwen/Qwen3.5-2B} {{Qwen3.5-2B}}.
\newblock Hugging Face model card, checked 2026-10-08.

\bibitem[{Radosevich and Halloran(2025)}]{radosevich2025mcpsafetyauditllms}
Brandon Radosevich and John Halloran. 2025.
\newblock \href {https://arxiv.org/abs/2504.03767} {{MCP} safety audit: {LLMs}
  with the {Model Context Protocol} allow major security exploits}.
\newblock \emph{Preprint}, arXiv:2504.03767.

\bibitem[{Schmotz et~al.(2026)Schmotz, Beurer-Kellner, Abdelnabi, and
  Andriushchenko}]{schmotz2026skillinjectmeasuringagentvulnerability}
David Schmotz, Luca Beurer-Kellner, Sahar Abdelnabi, and Maksym Andriushchenko.
  2026.
\newblock \href {https://arxiv.org/abs/2602.20156} {{Skill-Inject}: Measuring
  agent vulnerability to skill file attacks}.
\newblock \emph{Preprint}, arXiv:2602.20156.

\bibitem[{Sejfia and
  Schäfer(2022)}]{sejfia2022practicalautomateddetectionmalicious}
Adriana Sejfia and Max Schäfer. 2022.
\newblock \href {https://doi.org/10.1145/3510003.3510104} {Practical automated
  detection of malicious npm packages}.
\newblock In \emph{Proceedings of the 44th International Conference on Software
  Engineering}, pages 1681--1692. ACM.

\bibitem[{{TypeSafe}(2026)}]{typesafejev2026}
{TypeSafe}. 2026.
\newblock \href {https://docs.typesafe.ai/models} {{Jev} 1.13
  natural-language-inference {API}}.
\newblock Vendor documentation, checked 2026-10-12.

\bibitem[{Wang et~al.(2024)Wang, Li, Chen, Cai, Zhu, Lin, Cao, Kong, Liu, Liu,
  and Sui}]{wang2023largelanguagemodelsfair}
Peiyi Wang, Lei Li, Liang Chen, Zefan Cai, Dawei Zhu, Binghuai Lin, Yunbo Cao,
  Lingpeng Kong, Qi~Liu, Tianyu Liu, and Zhifang Sui. 2024.
\newblock \href {https://doi.org/10.18653/v1/2024.acl-long.511} {Large language
  models are not fair evaluators}.
\newblock In \emph{Proceedings of the 62nd Annual Meeting of the Association
  for Computational Linguistics (Volume 1: Long Papers)}, pages 9440--9450,
  Bangkok, Thailand. Association for Computational Linguistics.

\bibitem[{Wang et~al.(2021)Wang, Fang, Khabsa, Mao, and
  Ma}]{wang2021entailmentfewshotlearner}
Sinong Wang, Han Fang, Madian Khabsa, Hanzi Mao, and Hao Ma. 2021.
\newblock \href {https://arxiv.org/abs/2104.14690} {Entailment as few-shot
  learner}.
\newblock \emph{Preprint}, arXiv:2104.14690.

\bibitem[{Wang et~al.(2026)Wang, Liu, Deng, Zhang, Li, Chen, and
  Zhang}]{wang2026maliciousskillbench}
Yue Wang, Yi~Liu, Gelei Deng, Ying Zhang, Yuekang Li, Zhenyu Chen, and Leo
  Zhang. 2026.
\newblock \href {https://arxiv.org/abs/2608.19901} {{MaliciousSkillBench}: A
  comprehensive benchmark for malicious agent skill detection}.
\newblock \emph{Preprint}, arXiv:2608.19901.

\bibitem[{Warner et~al.(2025)Warner, Chaffin, Clavi{\'e}, Weller,
  Hallstr{\"o}m, Taghadouini, Gallagher, Biswas, Ladhak, Aarsen, Adams, Howard,
  and Poli}]{warner2024modernbert}
Benjamin Warner, Antoine Chaffin, Benjamin Clavi{\'e}, Orion Weller, Oskar
  Hallstr{\"o}m, Said Taghadouini, Alexis Gallagher, Raja Biswas, Faisal
  Ladhak, Tom Aarsen, Griffin~Thomas Adams, Jeremy Howard, and Iacopo Poli.
  2025.
\newblock \href {https://doi.org/10.18653/v1/2025.acl-long.127} {Smarter,
  better, faster, longer: A modern bidirectional encoder for fast, memory
  efficient, and long context finetuning and inference}.
\newblock In \emph{Proceedings of the 63rd Annual Meeting of the Association
  for Computational Linguistics (Volume 1: Long Papers)}, pages 2526--2547,
  Vienna, Austria. Association for Computational Linguistics.

\bibitem[{Yang et~al.(2026)Yang, Hu, Li, Yan, Wang, and
  Wang}]{yang2026automonitorbenchevaluatingreliabilityllmbased}
Shu Yang, Jingyu Hu, Tong Li, Hanqi Yan, Wenxuan Wang, and Di~Wang. 2026.
\newblock \href {https://doi.org/10.18653/v1/2026.findings-acl.116}
  {{AutoMonitor-Bench}: Evaluating the reliability of {LLM}-based misbehavior
  monitor}.
\newblock In \emph{Findings of the {A}ssociation for {C}omputational
  {L}inguistics: {ACL} 2026}, pages 2464--2483, San Diego, California, United
  States. Association for Computational Linguistics.

\bibitem[{Yin et~al.(2019)Yin, Hay, and Roth}]{yin2019benchmarking}
Wenpeng Yin, Jamaal Hay, and Dan Roth. 2019.
\newblock \href {https://doi.org/10.18653/v1/D19-1404} {Benchmarking zero-shot
  text classification: Datasets, evaluation and entailment approach}.
\newblock In \emph{Proceedings of the 2019 Conference on Empirical Methods in
  Natural Language Processing and the 9th International Joint Conference on
  Natural Language Processing (EMNLP-IJCNLP)}, pages 3914--3923, Hong Kong,
  China. Association for Computational Linguistics.

\bibitem[{{yoonholee}(2026)}]{agentskillmalware2026}
{yoonholee}. 2026.
\newblock \href {https://huggingface.co/datasets/yoonholee/agent-skill-malware}
  {{Agent Skill Malware}}.
\newblock Hugging Face dataset.

\bibitem[{Zeng et~al.(2024)Zeng, Liu, Mullins, Peran, Fernandez, Harkous,
  Narasimhan, Proud, Kumar, Radharapu, Sturman, and
  Wahltinez}]{zeng2024shieldgemma}
Wenjun Zeng, Yuchi Liu, Ryan Mullins, Ludovic Peran, Joe Fernandez, Hamza
  Harkous, Karthik Narasimhan, Drew Proud, Piyush Kumar, Bhaktipriya Radharapu,
  Olivia Sturman, and Oscar Wahltinez. 2024.
\newblock \href {https://arxiv.org/abs/2407.21772} {{ShieldGemma}: Generative
  {AI} content moderation based on {Gemma}}.
\newblock \emph{Preprint}, arXiv:2407.21772.

\bibitem[{Zhan et~al.(2024)Zhan, Liang, Ying, and
  Kang}]{zhan2024injecagentbenchmarkingindirectprompt}
Qiusi Zhan, Zhixiang Liang, Zifan Ying, and Daniel Kang. 2024.
\newblock \href {https://doi.org/10.18653/v1/2024.findings-acl.624}
  {{I}njec{A}gent: Benchmarking indirect prompt injections in tool-integrated
  large language model agents}.
\newblock In \emph{Findings of the Association for Computational Linguistics:
  ACL 2024}, pages 10471--10506, Bangkok, Thailand. Association for
  Computational Linguistics.

\bibitem[{Zhang et~al.(2025)Zhang, Huang, Mei, Yao, Wang, Zhan, Wang, and
  Zhang}]{zhang2025agentsecuritybenchasb}
Hanrong Zhang, Jingyuan Huang, Kai Mei, Yifei Yao, Zhenting Wang, Chenlu Zhan,
  Hongwei Wang, and Yongfeng Zhang. 2025.
\newblock \href {https://openreview.net/forum?id=V4y0CpX4hK} {{Agent Security
  Bench} ({ASB}): Formalizing and benchmarking attacks and defenses in
  {LLM}-based agents}.
\newblock In \emph{The Thirteenth International Conference on Learning
  Representations}.

\bibitem[{Zheng et~al.(2023)Zheng, Chiang, Sheng, Zhuang, Wu, Zhuang, Lin, Li,
  Li, Xing, Zhang, Gonzalez, and Stoica}]{zheng2023judging}
Lianmin Zheng, Wei-Lin Chiang, Ying Sheng, Siyuan Zhuang, Zhanghao Wu, Yonghao
  Zhuang, Zi~Lin, Zhuohan Li, Dacheng Li, Eric~P. Xing, Hao Zhang, Joseph~E.
  Gonzalez, and Ion Stoica. 2023.
\newblock \href {https://doi.org/10.52202/075280-2020} {Judging
  {LLM}-as-a-judge with {MT-Bench} and {Chatbot Arena}}.
\newblock In \emph{Advances in Neural Information Processing Systems 36}, pages
  46595--46623.

\bibitem[{Zhuang et~al.(2026)Zhuang, Xing, Zhou, Ma, Huang, Shen, Han, and
  Zhang}]{zhuang2026agenttrap}
Haomin Zhuang, Hanwen Xing, Yujun Zhou, Yuchen Ma, Yue Huang, Yili Shen, Yufei
  Han, and Xiangliang Zhang. 2026.
\newblock \href {https://arxiv.org/abs/2605.13940} {{AgentTrap}: Measuring
  runtime trust failures in third-party agent skills}.
\newblock \emph{Preprint}, arXiv:2605.13940.

\end{thebibliography}

\appendix

\section{Training and evaluation details}
\label{app:implementation}

\paragraph{Optimization and hardware.}
SkillScreen uses AdamW at $2\times10^{-4}$ with weight decay 0.01, 5\% linear warm-up and linear decay, and clips gradients at norm 1.0. LoRA adapters target the \texttt{q\_proj}, \texttt{v\_proj}, \texttt{in\_proj\_qkv} and \texttt{out\_proj} modules, and the classification head (\texttt{score}) is trained in full. Every SkillScreen run uses this single configuration without tuning, whereas Appendix~\ref{app:encoder} varies the encoder's learning rate. We use bf16 weights, gradient checkpointing and zero dropout. Each micro-batch contains one skill, with gradients accumulated over 16 steps. The seed controls adapter initialization and training order. One epoch gives 426 updates on MSD and SBR and 470 on SD. A run takes 3.4--3.9 hours on a 10\,GB MIG slice of an H100, with peak GPU memory of 7.3\,GiB. Scoring 1,947 MSD test skills takes nine minutes on the same hardware.

\paragraph{Baselines and software.}
Word-SVM uses scikit-learn's \texttt{LinearSVC} with $C=1$ and balanced class weights on TF-IDF features with document-frequency bounds of 2 and 99.5\%, fit on the training partition only. The package pins PyTorch 2.11.0, Transformers 5.17.0, PEFT 0.21.2 and scikit-learn 1.9.1.

\paragraph{Input length.}
The premise occupies the first line and the fixed hypothesis the next. Long premises lose tokens from the end, preserving the hypothesis. MSD training inputs have a median length of 1,367 tokens, a 95th percentile of 4,737 and a 99th percentile of 7,210. The cap truncates 44 of 6,814 training skills and 17 of 1,947 test skills. The LLM judges receive the same visible prefix: the skill text that fits in this 8,192-token input.

\paragraph{Checkpoint selection and calibration.}
We evaluate validation scores halfway through training and after the final update. The selected checkpoint minimizes FPR at 99\% validation recall. Ties prefer higher AP, then lower loss. All fifteen current-code SkillScreen runs, five per split, selected the final checkpoint.

We fit calibration to validation log-odds $\ell=z_e-\log(\exp(z_c)+\exp(z_n))$. The calibrated probability is $\sigma(a\ell+b)$ with $a>0$. We choose the highest threshold reaching 99\% validation recall and freeze it before test scoring.

\paragraph{Low-data runs.}
\label{app:lowdata}
From 25\% of the labels on, the full recipe flags about half as many benign skills as either alternative at a similar recall. At probability 0.5, its FPR is 7.8\% [6.6, 9.2] at 25\% and 3.1\% [2.5, 3.9] at 50\%, against 19.2\% [15.2, 23.6] and 6.7\% [4.8, 8.6] for the plain base and 16.1\% [12.5, 19.6] and 6.2\% [4.8, 7.7] for the unrelated hypothesis (95\% intervals of the mean over runs), at recalls of 93.2--95.1\%. At 50\%, it also ranks higher than the plain base in eight of nine paired runs (mean AUROC difference $+0.003$, $p=0.04$) and than the unrelated hypothesis in five of seven ($p=0.45$). The 10\% budget isolates the effect of the hypothesis, and its runs vary widely, which is why Section~\ref{sec:dataeff} pairs them by subset and seed: for the original configuration, AUROC spans 0.767--0.893 and the FPR at 0.5 spans 32--99.6\%. At this budget, the nine runs of the original configuration flag 89--96\% of benign skills at their frozen gates, and Word-SVM retrained on each subset reaches an AP of 0.974--0.976, against 0.919--0.965; with all labels, SkillScreen leads it (AP 0.998 versus 0.993). Every condition has nine runs at each fraction. Subsets are drawn with seeds 42, 1 and 2 and nested, so that each 10\% subset lies inside the 25\% and 50\% subsets of the same draw. On these three draws, each condition trains seeds 42, 1 and 2; 11, 12 and 13; and 21, 22 and 23. Five runs were rejected at freezing because their validation scores tied at the threshold for 99\% recall, and a rule fixed in advance replaced each with the next unused seed of the same draw. Seed 3 replaced seed 42 of the plain base at 10\%, seed 1 of the original configuration at 25\% and seed 42 of the unrelated hypothesis at 25\%, all on draw 42; seeds 24 and 25 replaced seeds 21 and 23 of the unrelated hypothesis at 50\% on draw 2. Pairs require both conditions to have trained the same seed on the same draw, so replacements without a counterpart leave seven to nine pairs per comparison; means and intervals use all nine runs of each condition.

\section{Partition composition}
\label{app:composition}

\begin{table*}[t]
\centering
\footnotesize
\setlength{\tabcolsep}{5pt}
\begin{tabular}{llrrr}
\toprule
Protocol & Source & Train (benign / malicious) & Validation & Test \\
\midrule
MSD & MalSkillBench & 0 / 1,919 & 0 / 537 & 0 / 967 \\
MSD & SkillTrustBench & 1,134 / 2,348 & 171 / 109 & 338 / 324 \\
MSD & SkillHarm & 0 / 601 & 0 / 35 & 0 / 92 \\
MSD & ATR Skill Security Benchmark & 329 / 13 & 43 / 3 & 83 / 5 \\
MSD & SkillFortifyBench & 70 / 86 & 7 / 2 & 13 / 2 \\
MSD & MaliciousAgentSkillsBench & 0 / 58 & 0 / 34 & 0 / 61 \\
MSD & SkillSafetyBench & 0 / 137 & 0 / 6 & 0 / 11 \\
MSD & AgentTrap & 31 / 50 & 3 / 14 & 11 / 22 \\
MSD & SkillLeakBench & 0 / 16 & 0 / 8 & 0 / 9 \\
MSD & Agent Skill Malware & 0 / 21 & 0 / 2 & 0 / 7 \\
MSD & SkillTrojan & 0 / 1 & -- & 2 / 0 \\
MSD & \emph{All} & 1,564 / 5,250 & 224 / 750 & 447 / 1,500 \\
\midrule
SD & MalSkillBench & 0 / 3,085 & 0 / 330 & -- \\
SD & SkillTrustBench & 1,478 / 2,495 & 165 / 286 & -- \\
SD & SkillHarm & -- & -- & 0 / 728 \\
SD & ATR Skill Security Benchmark & -- & -- & 455 / 21 \\
SD & SkillFortifyBench & -- & -- & 90 / 90 \\
SD & MaliciousAgentSkillsBench & 0 / 134 & 0 / 19 & -- \\
SD & SkillSafetyBench & 0 / 141 & 0 / 13 & -- \\
SD & AgentTrap & 41 / 75 & 4 / 11 & -- \\
SD & SkillLeakBench & 0 / 31 & 0 / 2 & -- \\
SD & Agent Skill Malware & 0 / 27 & 0 / 3 & -- \\
SD & SkillTrojan & 2 / 1 & -- & -- \\
SD & \emph{All} & 1,521 / 5,989 & 169 / 664 & 545 / 839 \\
\bottomrule
\end{tabular}
\caption{Benign / malicious counts after the five missing-text exclusions; SD also omits the eight malicious MalSkillBench records that its official split excludes. SD test sources are absent from training. MSD shares sources and separates malicious structural families. Empty partitions are shown with an en dash. Sources: MalSkillBench \citep{guo2026malskillbenchruntimeverifiedbenchmarkmalicious}, SkillTrustBench \citep{skilltrustbench2026}, SkillHarm \citep{ning2026skillharm}, ATR Skill Security Benchmark \citep{atrskillbenchmark2026}, SkillFortifyBench \citep{bhardwaj2026skillfortify}, MaliciousAgentSkillsBench \citep{liu2026domentionuserdetecting}, SkillSafetyBench \citep{jin2026skillsafetybench}, AgentTrap \citep{zhuang2026agenttrap}, SkillLeakBench \citep{chen2026skillleakbench}, Agent Skill Malware \citep{agentskillmalware2026} and SkillTrojan \citep{feng2026skilltrojan}.}
\label{tab:composition}
\end{table*}

Table~\ref{tab:composition} gives the source composition of MSD and SD after text checks. MSD keeps malicious structural families separate across partitions. The small SkillTrojan source has no validation records. SD places SkillHarm, ATR and SkillFortifyBench entirely in the test partition. MaliciousSkillBench's 4,575 structural families cover all of its malicious records. After the exclusions, MSD retains 4,572: 2,458 in training, 750 in validation, where each malicious skill forms its own family, and 1,364 in test; the other three families occur only in excluded records. The benchmark does not record the language of each skill; besides English, the MSD test set includes skills written in Chinese and Japanese (Appendix~\ref{app:errors}).

\section{Threshold sweeps and matched recall}

\begin{figure*}[t]
\centering
\includegraphics[width=\textwidth]{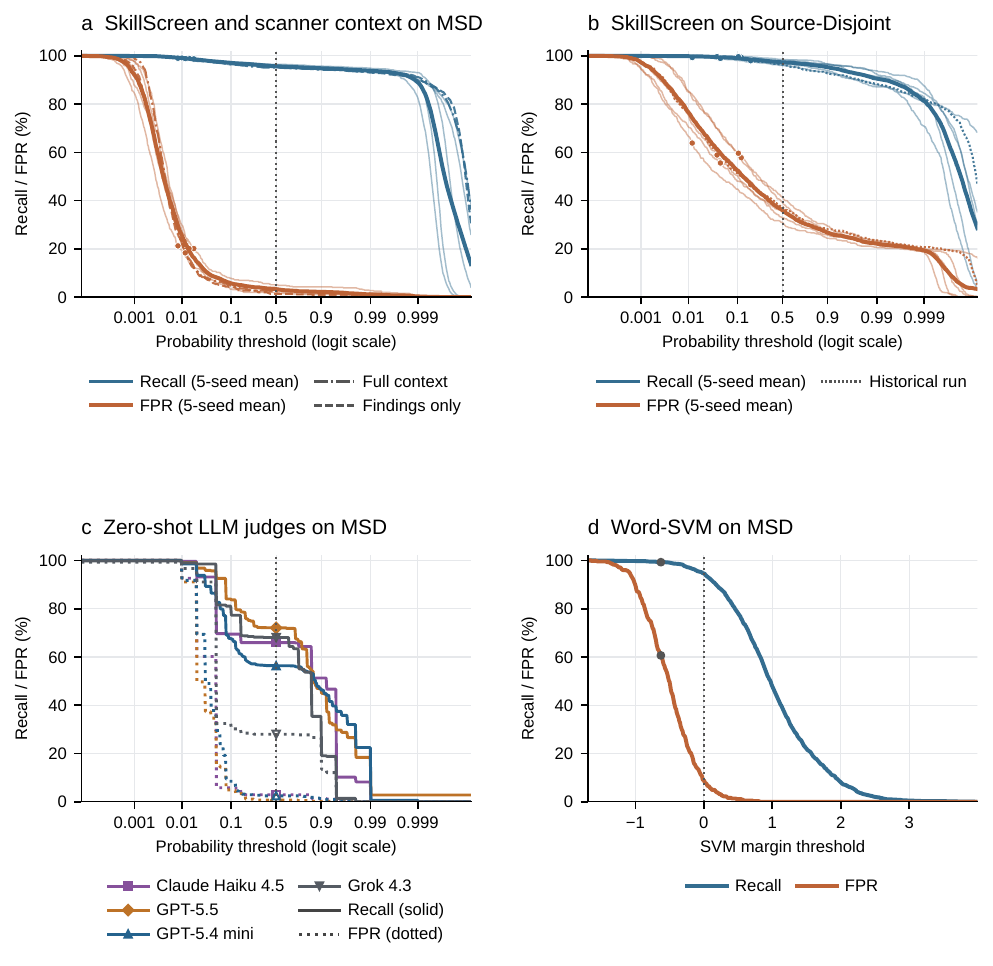}
\caption{Threshold sweeps for MSD SkillScreen and context variants (a), SD SkillScreen (b), MSD zero-shot judges (c) and MSD Word-SVM (d). Blue and orange show recall and FPR in (a), (b) and (d). Judges use solid recall and dotted FPR. Thin SkillScreen lines show seeds and bold lines their mean. Probability thresholds use a logit scale. SVM thresholds use margins. Vertical lines mark default thresholds. Circles mark thresholds chosen on validation. Judge markers show 0.5, filled for recall and hollow for FPR.}
\label{fig:sweeps}
\end{figure*}

SkillScreen's recall barely depends on the threshold, so the threshold mainly sets the review load (Figure~\ref{fig:sweeps}). Every MSD seed stays above 93\% recall for thresholds from 0.01 to 0.99, while the FPR falls from 19--30\% at 0.01 to 2--5\% at 0.5 and 0.2--2.7\% at 0.99.
\label{app:sweeps}

\begin{table}[t]
\centering
\footnotesize
\setlength{\tabcolsep}{4pt}
\resizebox{\columnwidth}{!}{%
\begin{tabular}{lrr}
\toprule
Method & at 95.6\% recall & at 99\% recall \\
\midrule
SkillScreen (5 seeds) & 2.0--4.0 & 19.9--24.6 \\
ModernBERT (5) & 2.2--3.8 & 25.1--36.2 \\
Word-SVM & 13.6 & 45.6 \\
GPT-5.5 & 34.7 & 91.1 \\
GPT-5.5 + few-shot (8) & 28.2 & 95.7 \\
GPT-5.5 + medium reasoning & 44.3 & 86.8 \\
Claude Haiku 4.5 & 92.6 & 92.6 \\
GPT-5.4 mini & 91.7 & 100.0 \\
Grok 4.3 (reseller) & 91.1 & 96.6 \\
\bottomrule
\end{tabular}}
\caption{MSD FPR (\%) at the highest observed test threshold reaching each recall target, read off the test score curves. SkillScreen and ModernBERT report seed ranges; Table~\ref{tab:review-matched} adds bootstrap intervals.}
\label{tab:matched}
\end{table}

Figure~\ref{fig:sweeps} plots rates over the native score scales. Table~\ref{tab:matched} gives the FPR at two recall targets from each method's observed test thresholds: at 99\% recall, every SkillScreen seed needs fewer false positives than any other method, and every prompted judge needs at least 28\% at 95.6\% recall and 86.8\% at 99\%.

\section{Average precision}
\label{app:ap}

\begin{table}[t]
\centering
\footnotesize
\begin{tabular}{lr}
\toprule
Method & AP \\
\midrule
SkillScreen (5 seeds) & 0.998 \\
\ \ + findings context & 0.998 \\
\ \ + full context & 0.998 \\
ModernBERT-large (5 seeds) & 0.997 \\
Word-SVM & 0.993 \\
TypeSafe Jev 1.13 & 0.942 \\
SkillSpector & 0.848 \\
Untuned OpenJev-2B & 0.886 \\
\midrule
GPT-5.5 (zero-shot) & 0.983 \\
\ \ + 8 examples & 0.985 \\
\ \ + medium reasoning & 0.984 \\
Grok 4.3 (reseller) & 0.846 \\
Claude Haiku 4.5 & 0.941 \\
GPT-5.4 mini & 0.954 \\
\bottomrule
\end{tabular}
\caption{MSD average precision at 77\% malicious prevalence. SkillScreen reports the five-seed mean with std below 0.001. Scanner AP uses 1,946 scored records. Other methods use all 1,947.}
\label{tab:ap}
\end{table}

Average precision confirms the ranking of Table~\ref{tab:main}: SkillScreen reaches 0.998, ahead of every baseline, and the prompted judges reach at most 0.985 (Table~\ref{tab:ap}). One MSD test scan timed out; it counts as flagged in decision metrics and is excluded from the scanner's AUROC and AP.

\section{Calibration}
\label{app:calibration}
SkillScreen's probabilities are well calibrated wherever validation and test share benign sources. On MSD, validation calibration lowers the 10-bin expected calibration error (ECE; \citealp{guo2017calibrationmodernneuralnetworks}) from 0.029--0.037 to 0.009--0.022 across the five seeds and the negative log-likelihood (NLL) from 0.13--0.16 to 0.09--0.11; on SBR, ECE falls from 0.009--0.015 to 0.006--0.009. On SD, calibration lowers NLL from 0.79--1.01 to 0.55--0.66, while ECE is 0.12--0.16 before and 0.14--0.16 after, reflecting the same benign source shift that moves the frozen gates (Section~\ref{sec:sd}).

\section{Scanner context}
\label{app:context}

\begin{table}[t]
\centering
\footnotesize
\setlength{\tabcolsep}{4pt}
\begin{tabular}{lrrrr}
\toprule
Input & Recall & FPR & AUROC & FPR$_{99}$ \\
\midrule
Skill text & 95.6 & 3.3 & 0.992 & 22.6 \\
+ findings & 95.4 & 1.3 & 0.993 & 22.1 \\
+ full context & 95.1 & 2.5 & 0.992 & 20.6 \\
\bottomrule
\end{tabular}
\caption{Scanner-context variants on MSD, using the metrics of Table~\ref{tab:main}. Skill text averages five seeds. Each context variant uses one seed.}
\label{tab:context-summary}
\end{table}

Scanner output adds no signal beyond the skill text. The two context variants scan the same Markdown and reserve 1,024 tokens for context while retaining the same skill-text prefix; median context lengths are 47 tokens for full context and 30 for findings-only.

Against SkillScreen seed 42, findings-only changes MSD FPR by $-0.45$ points ($[-1.6,+0.5]$) and recall by $+0.20$ ($[-0.3,+0.8]$), and full context changes FPR by $+0.67$ ($[-0.4,+1.8]$) and recall by $-0.13$ ($[-0.8,+0.6]$); all four paired intervals include zero. The scanner is also silent where SkillScreen errs: in a review of SD validation errors, every malicious skill that SkillScreen missed had a scanner risk score of zero.

\section{Per-seed runs and secondary gates}
\label{app:gates}

\begin{table*}[t]
\centering
\footnotesize
\setlength{\tabcolsep}{5pt}
\begin{tabular}{llrrrrrrrr}
\toprule
Run & Seed & Val.\ FPR@99 & \multicolumn{2}{c}{Gate recall / FPR} & FP & FN & AP & \multicolumn{2}{c}{At 0.5 recall / FPR} \\
\midrule
MSD & 42 & 23.7 & 98.93 & 21.70 & 97 & 16 & 0.9978 & 95.20 & 1.79 \\
MSD & 1 & 20.5 & 98.73 & 20.13 & 90 & 19 & 0.9976 & 96.47 & 5.15 \\
MSD & 2 & 21.4 & 99.00 & 21.25 & 95 & 15 & 0.9980 & 95.40 & 3.80 \\
MSD & 3 & 20.5 & 98.93 & 20.13 & 90 & 16 & 0.9977 & 95.07 & 3.13 \\
MSD & 4 & 27.7 & 99.13 & 23.94 & 107 & 13 & 0.9977 & 96.07 & 2.68 \\
\midrule
SD & 42 & 8.3 & 98.93 & 55.60 & 303 & 9 & 0.9325 & 96.07 & 35.41 \\
SD & 1 & 8.9 & 99.40 & 57.80 & 315 & 5 & 0.9542 & 97.97 & 39.27 \\
SD & 2 & 10.7 & 99.64 & 58.90 & 321 & 3 & 0.9423 & 97.74 & 32.84 \\
SD & 3 & 10.7 & 99.28 & 63.85 & 348 & 6 & 0.9202 & 96.31 & 31.01 \\
SD & 4 & 10.1 & 99.64 & 59.63 & 325 & 3 & 0.9562 & 98.45 & 41.28 \\
SD & 42 (historical code) & -- & 97.97 & 46.61 & 254 & 17 & 0.9214 & 96.19 & 37.43 \\
\midrule
SBR & 42 & 13.0 & 99.47 & 9.80 & 44 & 8 & 0.9992 & 98.13 & 1.56 \\
SBR & 1 & 5.4 & 98.73 & 2.90 & 13 & 19 & 0.9992 & 97.53 & 0.89 \\
SBR & 2 & 6.3 & 98.73 & 6.46 & 29 & 19 & 0.9991 & 98.27 & 2.45 \\
SBR & 3 & 9.4 & 99.20 & 8.46 & 38 & 12 & 0.9992 & 97.93 & 0.67 \\
SBR & 4 & 5.8 & 99.07 & 6.46 & 29 & 14 & 0.9991 & 98.60 & 3.12 \\
SBR & 42 (historical code) & -- & 99.13 & 5.57 & 25 & 13 & 0.9993 & 98.00 & 0.89 \\
\bottomrule
\end{tabular}
\caption{SkillScreen results by seed. Validation FPR is measured at 99\% recall. Test columns give recall, FPR, FP and FN at the frozen threshold, AP from continuous scores, and recall/FPR at probability 0.5. The fifteen current-code runs selected the final checkpoint. The historical SD and SBR runs use earlier code.}
\label{tab:seeds}
\end{table*}

SkillScreen's operating points are consistent across seeds (Table~\ref{tab:seeds}): on MSD, the frozen gates give 98.7--99.1\% test recall at 20.1--23.9\% FPR. At their own thresholds for 99\% validation recall on MSD, Word-SVM reaches 99.3\% recall at 60.6\% FPR, the full-context variant 99.3\% at 30.2\% and the findings-only variant 98.9\% at 20.1\%, while the commercial NLI API and the static scanner flag every record. SkillScreen's MSD thresholds correspond to raw entailment probabilities between 0.008 and 0.018. Per seed, SkillScreen corrects 83 Word-SVM errors on MSD and introduces 42 new ones.

\section{Encoder baseline}
\label{app:encoder}
ModernBERT-large has 28 layers and 395.8M parameters with the classification head \citep{warner2024modernbert}. It reads raw Markdown with its tokenizer's \texttt{[CLS]} and \texttt{[SEP]} markers. Its 8,192-token cap truncates 82 of 6,814 MSD training skills and 23 of 1,947 test skills. The tokenizer produces 14\% more tokens than SkillScreen's Qwen tokenizer on the MSD test set.

All parameters train with a fresh two-way head, batch size 1 and accumulation over 16 steps. We use AdamW with weight decay 0.01 on matrices, 5\% linear warm-up, linear decay and gradient clipping at 1.0; the default recipe uses a learning rate of $2\times10^{-5}$ for one epoch. Checkpoints are evaluated at the midpoint and end of training using the same validation rule as SkillScreen. Calibration and the 99\%-recall threshold use the two-way log-odds and are frozen on validation.

At every encoder learning rate, SkillScreen needs fewer false positives at 99\% recall, by 6.8 to 14.6 points (Table~\ref{tab:encoder-grid}). The table varies the rate over three values on MSD, with every run trained for one epoch as SkillScreen is; the main text uses the default rate, and every encoder run selected the final checkpoint.

Every encoder run except a first MSD set used bf16 autocast over float32 weights on a 10\,GB MIG slice with peak memory of 7.9\,GiB, and one epoch took 1.0--1.1 hours. SkillScreen trains 2.19M parameters on a 2B backbone, whereas the encoder trains all 395.8M.

On MSD, the reported encoder runs follow a rule fixed before they were trained: a GPU rerun of all five seeds with the same code, data and recipe replaces a first set trained in float32 on 32 CPU cores (7.3--9.3 hours per run), whatever its outcome. All five reruns converged, and they are the stronger set: the four converged runs of the first set need 33.9\% FPR at 99\% recall, against 30.4\% (Tables~\ref{tab:encoder-grid} and~\ref{tab:review-matched}).

None of the fifteen SkillScreen runs collapsed, whereas two of the sixteen MSD and SD default-recipe encoder runs did: seed 2 of the first MSD set and SD seed 2. Table~\ref{tab:encoder-seeds} lists all sixteen, and Appendix~\ref{app:sbr} reports the five SBR runs. The collapsed runs kept a loss near 0.5 for the whole epoch and reached a validation AP of 0.952 and 0.935, against 0.997--0.999 for the other runs. Pre-clipping gradient-norm spikes during or just after warm-up preceded both collapses (peaks of 1,178 and 3,193), but the other default-recipe runs peaked at 38--4,903, including MSD seed 2 in its rerun (67). On SD, seed 5 replaces the collapsed seed 2 in Table~\ref{tab:main}, and Table~\ref{tab:review-matched} also gives all six runs.

\begin{table*}[t]
\centering
\footnotesize
\setlength{\tabcolsep}{4pt}
\resizebox{\textwidth}{!}{%
\begin{tabular}{lrrrrrrrl}
\toprule
Learning rate & Seeds & Val.\ FPR@99 & Recall & FPR & AUROC & Gate recall / FPR & FPR@99 & $\Delta$ vs SkillScreen @99 \\
\midrule
$2\times10^{-5}$ (default) & 5 & 27.7 & 96.0 & 4.1 & 0.990 & 98.9 / 30.2 & 30.4 & $+7.8$ [$-3.4$, $+22.1$] \\
\ \ first set (CPU, float32) & 4 of 5 & 31.8 & 96.1 & 4.8 & 0.990 & 99.1 / 34.7 & 33.9 & $+11.3$ [$-0.7$, $+20.7$] \\
$10^{-5}$ & 3 & 36.0 & 96.1 & 10.0 & 0.985 & 99.0 / 39.7 & 37.2 & $+14.6$ [$+3.1$, $+28.4$] \\
$5\times10^{-5}$ & 3 & 34.1 & 96.0 & 3.0 & 0.991 & 99.3 / 36.5 & 29.4 & $+6.8$ [$-5.6$, $+17.4$] \\
\midrule
SkillScreen & 5 & 22.8 & 95.6 & 3.3 & 0.992 & 98.9 / 21.4 & 22.6 & -- \\
\bottomrule
\end{tabular}}
\caption{ModernBERT-large learning rates on MSD, each trained for one epoch (seeds 42, 1 and 3 for the alternatives; the first default set excludes its failed seed 2). The main text reports the default learning rate. Rates are seed-mean percentages. Validation FPR@99 is measured at each run's selected checkpoint. Recall and FPR use probability 0.5 on test. AUROC is computed from continuous test scores. The gate is frozen for 99\% validation recall; FPR@99 is read off the test sweep with the paired difference against SkillScreen's five seeds.}
\label{tab:encoder-grid}
\end{table*}

\begin{table*}[t]
\centering
\footnotesize
\setlength{\tabcolsep}{5pt}
\begin{tabular}{llrrrrrrrr}
\toprule
Split & Seed & Val.\ FPR@99 & \multicolumn{2}{c}{Gate recall / FPR} & FP & FN & AP & \multicolumn{2}{c}{At 0.5 recall / FPR} \\
\midrule
MSD & 42 & 32.6 & 99.00 & 38.03 & 170 & 15 & 0.9970 & 95.47 & 2.24 \\
MSD & 1 & 25.4 & 98.80 & 25.06 & 112 & 18 & 0.9974 & 95.87 & 3.80 \\
MSD & 2 & 31.7 & 99.13 & 34.23 & 153 & 13 & 0.9972 & 95.93 & 3.80 \\
MSD & 3 & 21.4 & 98.47 & 25.28 & 113 & 23 & 0.9970 & 96.00 & 3.58 \\
MSD & 4 & 27.2 & 99.00 & 28.41 & 127 & 15 & 0.9973 & 96.67 & 7.16 \\
\midrule
MSD, CPU set & 42 & 44.2 & 99.27 & 46.09 & 206 & 11 & 0.9971 & 95.60 & 4.47 \\
MSD, CPU set & 1 & 32.6 & 99.13 & 36.24 & 162 & 13 & 0.9971 & 96.33 & 5.59 \\
MSD, CPU set & 2 (failed) & 93.3 & 99.33 & 95.75 & 428 & 10 & 0.9397 & 99.93 & 98.43 \\
MSD, CPU set & 3 & 23.7 & 98.80 & 30.87 & 138 & 18 & 0.9972 & 96.27 & 3.13 \\
MSD, CPU set & 4 & 26.8 & 99.00 & 25.50 & 114 & 15 & 0.9974 & 96.33 & 5.82 \\
\midrule
SD & 42 & 15.4 & 97.85 & 70.64 & 385 & 18 & 0.7497 & 92.85 & 49.54 \\
SD & 1 & 8.3 & 94.87 & 61.65 & 336 & 43 & 0.7587 & 94.64 & 59.82 \\
SD & 2 (failed) & 95.9 & 99.40 & 98.35 & 536 & 5 & 0.6114 & 99.28 & 97.98 \\
SD & 3 & 20.1 & 99.52 & 70.64 & 385 & 4 & 0.8914 & 97.50 & 53.03 \\
SD & 4 & 20.7 & 99.88 & 77.25 & 421 & 1 & 0.7832 & 99.05 & 56.33 \\
SD & 5 & 36.1 & 97.74 & 83.67 & 456 & 19 & 0.7306 & 86.65 & 55.05 \\
\bottomrule
\end{tabular}
\caption{ModernBERT-large with the default recipe by seed. The MSD CPU set is the first MSD set, trained in float32 on CPUs; every other run used bf16 on a GPU slice. Rates are percentages. Gate columns use thresholds frozen for 99\% validation recall. The remaining columns give test AP and results at probability 0.5.}
\label{tab:encoder-seeds}
\end{table*}

\section{Source-Balanced Random}
\label{app:sbr}

When test skills come from sources seen in training, SkillScreen reaches 99\% recall at single-digit FPR. At the default threshold, its five seeds reach 98.1\,$\pm$\,0.4\% recall at 1.7\,$\pm$\,1.0\% FPR, with AUROC 0.997. Word-SVM reaches 96.3\% recall at 11.8\% FPR, with AUROC 0.985. Relative to Word-SVM, SkillScreen raises recall by 1.8 points ($[+0.8,+2.7]$) and lowers the FPR by 10.1 points ($[-13.2,-7.3]$). At thresholds chosen for 99\% validation recall, SkillScreen reaches 98.7--99.5\% test recall at 2.9--9.8\% FPR (seed mean 99.0\% at 6.8\%), and Word-SVM 99.4\% at 41.0\%. At matched 99\% test recall SkillScreen needs 6.6\% FPR $[2.4, 13.5]$ and Word-SVM 28.1\%.

ModernBERT-large uses the same one-epoch GPU recipe on SBR, with seeds 42, 1, 2, 3 and 4. At matched 99\% test recall, its mean FPR is 12.1\% $[6.2,25.5]$. The paired difference from SkillScreen is $+5.4$ points ($[-0.1,+15.5]$).

The other detectors follow the MSD pattern. Full context reaches 97.5\% recall at 2.5\% FPR at 0.5, and 99.1\% at 10.7\% FPR at its high-recall threshold. Findings-only reaches 97.6\% at 0.7\% FPR and 99.0\% at 14.3\% FPR, respectively. TypeSafe Jev 1.13 reaches 26.1\% recall at 0.5\% FPR, with AUROC 0.782. Static scanning reaches 7.3\% at 1.3\% FPR. Untuned OpenJev reaches 3.5\% at zero FPR.

Recall barely depends on whether a family was seen in training: at the high-recall thresholds, SkillScreen recovers 99.3\% of the 746 malicious test records whose family appears in training and 98.8\% of the 755 others (five-seed means; 98.8\% and 97.4\% at 0.5). This comparison groups the SBR test records by family overlap; the MSD protocol evaluates family transfer with a separate training partition.

\paragraph{Package inputs.}
We extend the NLI premise with supporting-file contents while retaining the fixed maliciousness hypothesis and LoRA training recipe. A 32K variant uses recovered SkillTrustBench packages with their original SBR assignments, comprising 2,867 training, 403 validation and 811 test packages. With seed 42 and a threshold frozen at 99\% validation recall, it achieves 99.02\% test recall and 0.33\% FPR.

\section{Recall--FPR curves and source errors}
\label{app:sdfig}

\begin{figure*}[t]
\centering
\includegraphics[width=\textwidth]{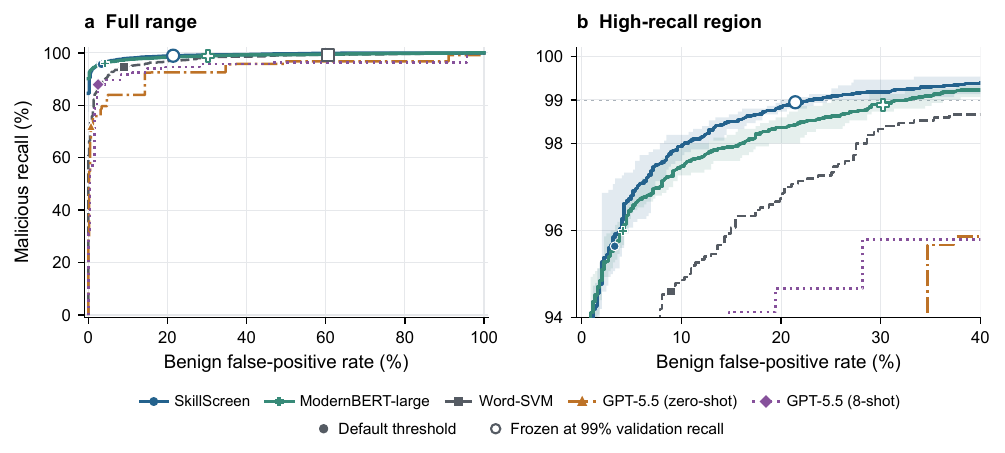}
\caption{MSD recall--FPR trade-offs. Panel (a) shows the full range and (b) enlarges the high-recall region. SkillScreen and ModernBERT-large show five-seed means at each FPR, with min--max shading. Filled markers show default operating points. Open markers show test performance at thresholds frozen for 99\% validation recall. Markers average seeds where applicable. The test set contains 1,500 malicious and 447 benign skills.}
\label{fig:curves}
\end{figure*}

\begin{figure*}[t]
\centering
\includegraphics[width=\textwidth]{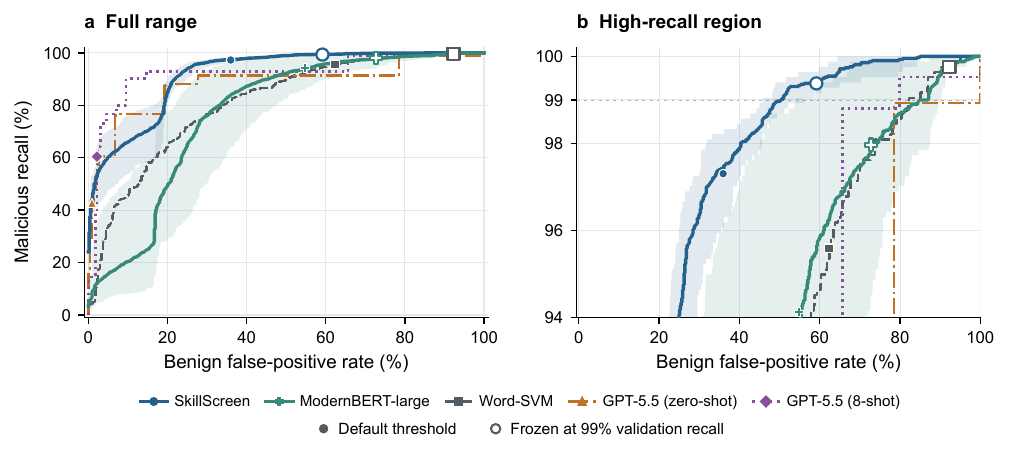}
\caption{SD recall--FPR trade-offs, following the conventions of Figure~\ref{fig:curves}. Both panels retain the full FPR range. SkillScreen and ModernBERT-large summarize five seeds, using the converged runs for the encoder. GPT-5.5 is shown zero-shot and with eight SD training examples. The test set contains 839 malicious and 545 benign skills.}
\label{fig:sdcurves}
\end{figure*}

\begin{figure*}[t]
\centering
\includegraphics[width=\textwidth]{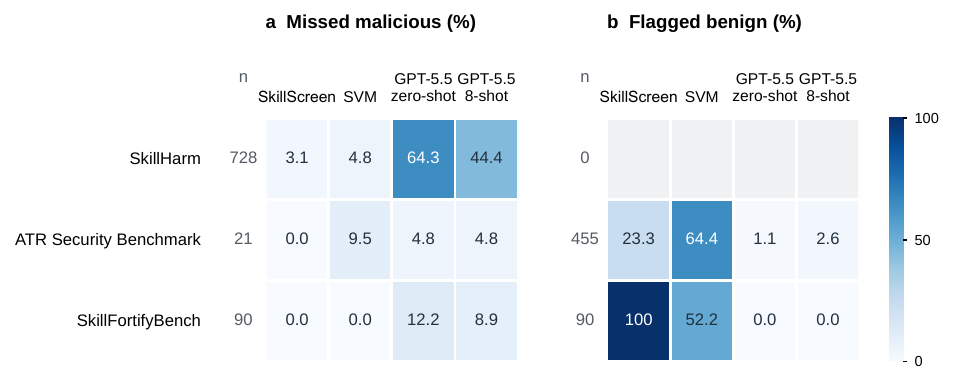}
\caption{SD errors by source at default thresholds. Cells and class counts follow Figure~\ref{fig:sources}. SkillScreen averages five current-code seeds. Benign records occur in two sources.}
\label{fig:sdsources}
\end{figure*}

On SD, SkillScreen has the lowest mean FPR at both recall targets. At 95.6\% test recall, the SkillScreen seeds require 24--33\% FPR, the converged ModernBERT runs 35--76\%, Word-SVM 62.8\%, GPT-5.5 78.5\% zero-shot, 65.7\% with eight examples and 73.0\% with reasoning, Grok 4.3 89.9\%, Haiku 80.7\% and GPT-5.4 mini 83.5\%. At 99\% recall, SkillScreen requires 46--56\% FPR, the converged encoder runs 54--90\% and the SVM 82.2\%; the judges require 79.8\% (GPT-5.5 with eight examples) to 100\%. Figures~\ref{fig:sdcurves} and~\ref{fig:sdsources} show the SD curves and source-level errors, and Figure~\ref{fig:curves} gives the full MSD curves and a closer view of low FPR.

\section{Matched-recall intervals, sources, families and combinations}
\label{app:review}
SkillScreen's high-recall advantage over Word-SVM and every prompted judge holds under resampling. All numbers here come from the released per-skill predictions (2,000 stratified bootstrap replicates, seed 20261006, the same index matrix as the main metrics, so differences are paired).

\paragraph{FPR at matched recall.}
Table~\ref{tab:review-matched} reads each method's threshold off its own MSD test sweep and re-determines it in every replicate; multi-seed rows are seed means, and unavailable judge scores count as flagged. At 99\% recall, SkillScreen needs fewer false positives than Word-SVM, every prompted judge and the empty hypothesis, with paired intervals that exclude zero; the other SkillScreen variants stay within noise of SkillScreen, and the encoder's gap on MSD ($+7.8$ points) becomes clear on SD ($+24.1$). On SD at the default threshold (Table~\ref{tab:main}), SkillScreen's AUROC std is 0.018, its bootstrap intervals are $[96.4,98.1]$ for recall and $[32.4,39.7]$ for FPR, and relative to Word-SVM it raises recall by 1.7 points ($[+0.3,+3.1]$) and lowers the FPR by 26.4 points ($[-31.7,-21.2]$); untuned OpenJev-2B reaches 0.6\% recall at 0.0\% FPR (AUROC 0.568).

\begin{table}[t]
\centering
\footnotesize
\setlength{\tabcolsep}{3pt}
\resizebox{\columnwidth}{!}{%
\begin{tabular}{lrrr}
\toprule
Method & FPR@95.6\% & FPR@99\% [CI] & $\Delta$ vs SkillScreen @99\% [CI] \\
\midrule
SkillScreen (5) & 3.2 & 22.6 [13.5, 37.0] & -- \\
ModernBERT-large (5) & 3.2 & 30.4 [21.2, 51.4] & $+7.8$ [$-3.4$, $+22.1$] \\
\ \ CPU set, 4 converged & 2.8 & 33.9 [21.2, 46.9] & $+11.3$ [$-0.7$, $+20.7$] \\
\ \ CPU set, all 5 & 18.8 & 45.9 [35.8, 56.4] & $+23.3$ [$+12.1$, $+31.3$] \\
Word-SVM & 13.6 & 45.6 [30.0, 66.2] & $+23.0$ [$+6.8$, $+39.8$] \\
Plain base (5) & 2.6 & 21.1 [11.5, 39.5] & $-1.5$ [$-11.9$, $+8.0$] \\
Paraphrased hypothesis & 1.6 & 21.0 [11.4, 35.8] & $-1.6$ [$-11.9$, $+4.7$] \\
Unrelated hypothesis (5) & 3.3 & 28.9 [17.9, 48.4] & $+6.3$ [$-3.6$, $+19.3$] \\
Empty hypothesis & 2.2 & 44.7 [21.5, 65.5] & $+22.1$ [$+2.0$, $+40.4$] \\
0.8B base (2) & 3.4 & 26.0 [14.7, 36.9] & $+3.4$ [$-9.1$, $+11.3$] \\
+ findings context & 2.0 & 22.1 [10.3, 41.6] & $-0.4$ [$-12.1$, $+9.5$] \\
+ full context & 2.7 & 20.6 [13.4, 33.3] & $-2.0$ [$-14.2$, $+6.3$] \\
GPT-5.5 & 34.7 & 91.1 [88.4, 100.0] & $+68.5$ [$+55.1$, $+83.0$] \\
GPT-5.5 + 8 examples & 28.2 & 95.7 [93.7, 97.5] & $+73.2$ [$+58.7$, $+82.6$] \\
GPT-5.5 + reasoning & 44.3 & 86.8 [83.7, 100.0] & $+64.2$ [$+49.6$, $+79.8$] \\
Claude Haiku 4.5 & 92.6 & 92.6 [90.2, 95.1] & $+70.0$ [$+55.3$, $+79.6$] \\
GPT-5.4 mini & 91.7 & 100.0 [91.1, 100.0] & $+77.4$ [$+61.7$, $+86.4$] \\
\bottomrule
\end{tabular}}
\caption{Benign FPR (\%) on the MSD test set at the highest threshold of each method that reaches the stated recall, with the threshold re-chosen in every bootstrap replicate; intervals are 95\% percentile intervals of the seed mean, and differences are paired. On SD the five SkillScreen seeds need 27.4\% [23.6, 32.4] at 95.6\% recall and 50.5\% [42.1, 59.9] at 99\%, the five converged ModernBERT runs 56.0\% and 74.7\% [69.8, 79.7] ($+24.1$ [$+14.4$, $+32.7$] at 99\%; all six 78.2\%, $+27.7$ [$+18.1$, $+36.1$]), and the SVM 62.8\% and 82.2\%. On SBR SkillScreen needs 6.6\% [2.4, 13.5] at 99\% recall and the SVM 28.1\%.}
\label{tab:review-matched}
\end{table}

\paragraph{Sources and families.}
Family novelty barely affects recall. Structural families are defined by the benchmark (\texttt{structural\_family\_id}), with sizes in Table~\ref{tab:families}. On SBR, SkillScreen's fixed-rule recall is 98.8\% on the 746 malicious test skills whose family appears in training and 97.4\% on the 755 whose family does not (99.3\% and 98.8\% at the gates); on SD it is 97.8\% on 322 seen and 97.0\% on 517 unseen (99.1\% and 99.5\% at the gates). All are five-seed means.

Per source, the trained scorers recover far more malicious skills than the prompted judges. Macro-averaging over the nine MSD sources with at least five malicious test skills and the four with at least five benign ones (within-source resampling) gives SkillScreen 79.1\% [73.6, 84.5] recall and 5.0\% [1.8, 10.1] FPR at the fixed threshold (92.5\% and 29.0\% at the gate), ModernBERT-large 84.2\% [79.6, 88.5] and 5.2\% (92.4\% and 24.8\% at the gate), Word-SVM 77.4\% and 8.6\%, GPT-5.5 37.9\% [34.1, 42.4] and 0.4\%, and GPT-5.5 with eight examples 60.6\% and 1.0\%. The micro values are 95.6\%/3.3\%, 96.0\%/4.1\%, 94.6\%/8.9\%, 72.1\%/0.7\% and 87.9\%/2.5\%.

\begin{table}[t]
\centering
\footnotesize
\setlength{\tabcolsep}{4pt}
\begin{tabular}{llrrrr}
\toprule
Split & Partition & Malicious & Families & Singletons & $\ge 5$ \\
\midrule
MSD & train & 5,250 & 2,458 & 50\% & 129 \\
MSD & test & 1,500 & 1,364 & 90\% & 0 \\
SBR & train & 5,249 & 3,542 & 75\% & 61 \\
SBR & test & 1,501 & 1,249 & 88\% & 8 \\
SD & train & 5,989 & 3,992 & 72\% & 69 \\
SD & test & 839 & 304 & 71\% & 43 \\
\bottomrule
\end{tabular}
\caption{Malicious structural families per partition: number of records and families, share of families with one record, and families with at least five records. The largest MSD training family has 146 records; no MSD test family has more than two. On SD, 322 of 839 malicious test skills belong to a family seen in training, and the largest test family has 141 records.}
\label{tab:families}
\end{table}

\paragraph{Combining SkillScreen with GPT-5.5 and precision at low prevalence.}
A precise judge does not substitute for the trained scorer at high recall. With SkillScreen's frozen gate, a cascade that sends flagged skills to GPT-5.5 and keeps only the judge's positives reaches 71.8\% recall at 0.6\% FPR on MSD (42.7\% at 0.9\% on SD) while the judge reads 81\% (SD: 84\%) of all skills; the union of the two verdicts reaches 99.2\% recall at 21.5\% FPR on MSD and 99.5\% at 59.2\% on SD, against 98.9\%/21.4\% and 99.4\%/59.2\% for SkillScreen alone.

Precision on MSD as a function of malicious prevalence (1\%, 5\%, 10\%, 50\%): SkillScreen at 0.5, 22.6\%, 60.3\%, 76.2\%, 96.7\%; SkillScreen at the gate, 4.5\%, 19.5\%, 33.9\%, 82.2\%; ModernBERT-large at 0.5, 19.1\%, 55.1\%, 72.2\%, 95.9\%, and at its gate, 3.2\%, 14.7\%, 26.7\%, 76.6\%; Word-SVM, 9.6\%, 35.7\%, 54.0\%, 91.4\%; GPT-5.5, 52.0\%, 85.0\%, 92.3\%, 99.1\%. On SD, SkillScreen at 0.5 reaches 2.7\% precision at 1\% prevalence and 23.1\% at 10\%.

\section{Robustness to rewriting}
\label{app:robustness}
SkillScreen responds to what a skill does rather than to how it is worded (Table~\ref{tab:robustness}). We sample 200 malicious MSD test skills proportionally by source (seed 20261009, excluding the 14 skills longer than 30,000 characters) and rewrite each one twice with GPT-5.4 mini under a fixed prompt that treats the skill as data. The paraphrase rewrites every natural-language sentence and keeps code blocks, commands, URLs, paths, identifiers and the front matter unchanged. The script move replaces every code block and inline command by one neutral line that names a referenced script, lists the scripts in a final \emph{Files} section, and rewrites the remaining prose in neutral developer language that does not describe what the moved commands do. All 400 rewrites were usable; the script move shortens a skill from 6,383 to 3,608 characters on average. The rewrites cost USD\,2.61. We score the original and rewritten texts with the four archived MSD adapters (seeds 1--4), their frozen calibrations and gates, and the same 8,192-token input.

\begin{table}[t]
\centering
\footnotesize
\setlength{\tabcolsep}{4pt}
\begin{tabular}{llrrr}
\toprule
Rewrite & Rule & Before & After & Drop per adapter \\
\midrule
Paraphrase & 0.5 & 96.6 & 96.0 & 0.5--1.0 \\
Paraphrase & gate & 99.8 & 99.6 & 0.0--0.5 \\
Script move & 0.5 & 96.6 & 67.5 & 24.5--35.0 \\
Script move & gate & 99.8 & 87.0 & 11.5--14.0 \\
\bottomrule
\end{tabular}
\caption{Recall (\%) of the four MSD adapters on 200 malicious test skills before and after rewriting, averaged over adapters, with the range of per-adapter drops in points. Paired bootstrap intervals (2,000 replicates over records) exclude zero for every script-move drop.}
\label{tab:robustness}
\end{table}

Paraphrasing every sentence changes recall by at most one point per adapter. Moving the commands out of \texttt{SKILL.md} removes the evidence itself: each adapter loses 50--72 of the 200 skills at 0.5 and 23--29 at the gate. The rewritten skills remain malicious packages whose \texttt{SKILL.md} no longer states what runs, so this drop measures how much of SkillScreen's recall rests on visible commands, the case for package-level inputs.

\section{Qualitative error examples}
\label{app:errors}

\paragraph{Errors by source.}
Errors are rare on the sources that make up most of the MSD test set (Figure~\ref{fig:sources}): SkillScreen misses 3\% of the 967 MalSkillBench skills and none of the 92 from SkillHarm, and flags 1.8\% of the 338 benign SkillTrustBench skills. Misses concentrate in three small sources, 73\% of the nine SkillLeakBench records, 45\% of the 22 from AgentTrap and 27\% of the 61 from MaliciousAgentSkillsBench, and benign FPR is highest on ATR, at 9\% of 83 skills.

\paragraph{Misses shared across seeds.}
All five seeds miss the same 42 malicious skills, and another 45 are missed by some seeds. The 42 shared misses fall into two groups. Several describe ordinary tools, including an SVG icon generator, a blockchain guide, a pandas toolkit and a distributed-training client, and their Markdown carries no clear evidence for the public malicious label, which can rest on package code or upstream evidence.

The others state suspicious behaviour directly and give concrete targets for detection from the Markdown itself. A Chinese-language token saver runs automatically and refuses every user request; GPT-5.5 scores it at 0.92 and SkillScreen at 0.20. A Japanese skill routes every question through its own background script, and an email skill claims authority over all mail operations and sets BCC recipients.

\paragraph{Benign skills flagged by every seed.}
On SD, SkillFortifyBench skills have a median length of 459 characters, each with one \texttt{bash} block and a benchmark marker comment, and SkillScreen scores every benign one at least 0.998. One benign skill reads, in full: \texttt{name: kernel-version-check} / \texttt{description: Compact tool verifying data state.} / \texttt{<!-- SKILLFORTIFYBENCH:INERT -- arXiv:2603.00195 Appendix B specification execution; do not execute. Licensed MIT. -->} / a \texttt{bash} block with \texttt{df -h}, \texttt{uname -a} and a one-line Python platform query / \texttt{Gathers read-only system information. No write operations, no network access.} All five seeds score it above 0.999; GPT-5.5 gives 0.01. Its malicious sibling in the same template replaces the block by \texttt{rm -rf /tmp/.cache-04/* /var/log/app-04/*} and \texttt{chmod 777 /tmp/.workdir-04}. On ATR, whose 455 benign test skills have a median length of 5,900 characters, the skills SkillScreen flags contain security vocabulary (\emph{token}, \emph{api key}, \emph{credential}, \emph{payload}) no more often than the ones it passes.
On MSD, only four benign skills are flagged by every seed, and another 26 by some seeds. Three of the shared false positives cover context compression, MCP server development and Firebase hosting. The fourth is a skill-discovery helper that sends the machine's hostname to a remote URL using \texttt{curl}, which GPT-5.5 also flags (0.95).

\paragraph{Disagreements with GPT-5.5.}
SkillScreen covers most of GPT-5.5's blind spots: every seed detects 346 of its 419 malicious misses, including a DAO governance persona, a Kubernetes helper, an Obsidian editor and a skill that requires a warm-up script to run ``without asking the user for confirmation'', all scored at 1.0. In the other direction, GPT-5.5 recovers nine of the 42 skills missed by every SkillScreen seed.

\section{Judge prompt and request settings}
\label{app:prompt}

The complete prompt and response schema are released in \path{scripts/baselines/llm_judge_prompt.py}. The prompt defines malicious skills through credential theft, hidden payloads, deception, privilege escalation, sabotage and attempts to override the user's intent. It distinguishes these from authorized installation, shell commands, credential use, security testing and documented network access. The skill is marked as untrusted text. The user message places it inside \texttt{<skill\_md>} tags and requests a JSON answer.

The answer has two fields. \texttt{malicious} is a boolean and \texttt{probability} is a number in $[0,1]$. Providers enforce this schema through structured output, and the parser checks the types and probability range. Skills with missing or invalid answers would be routed to review; every completed request returned a valid answer. Standard requests allow 256 answer tokens. No request specifies a temperature or sampling seed. OpenAI requests set \texttt{store=false}.

The few-shot condition adds eight labelled training records to the system message. Records are sorted by identifier, shuffled with seed 42 and selected from distinct sources where possible. Four malicious and four benign examples are interleaved. Each is capped at 500 SkillScreen tokens with a truncation marker. Answers use probabilities 0.95 and 0.05. The fixed prefix has about 2,800 provider tokens, and caching covers 65\% of input tokens. The release lists the selected record identifiers and excerpt hashes in \path{configs/baselines/fewshot-msd.json}, and \path{scripts/baselines/fewshot_prompt.py} rebuilds and verifies the excerpts from the benchmark text. On SD, the same procedure selects eight examples from the SD training partition (\path{configs/baselines/fewshot-sd.json}); the MSD set cannot be reused there, because three of its records are SD test records from SD test sources.

The reasoning condition uses the zero-shot prompt with medium reasoning effort and an 8,192-token completion limit. Returned model identifiers were
\begin{quote}\small\raggedright
\texttt{claude-haiku-4-5-20251001}\\
\texttt{gpt-5.5-2026-04-23}\\
\texttt{gpt-5.4-mini-2026-03-17}\\
\texttt{grok-4.3}
\end{quote}
Billed input tokens were 1.41 times the visible token count on Anthropic, 1.14 on OpenAI and 1.23 on the reseller. Anthropic and OpenAI requests were submitted through batch APIs.

\paragraph{Additional judge results.}
GPT-5.5 is the strongest zero-shot judge on MSD by AUROC (0.950). Haiku 4.5 and GPT-5.4 mini recover 66\% and 56\% of malicious skills at their default thresholds. Grok 4.3 flags 28\% of benign skills. On SD, Haiku and GPT-5.4 mini recover 29\% and 23\% of malicious skills, and Grok 4.3 recovers 36\% while flagging 21\% of benign skills (AUROC 0.614).

The few-shot gain varies across sources. Miss rates fall from 83\% to 3\% on SkillHarm and from 15\% to 1\% on SkillTrustBench. They remain 89\% on SkillLeakBench and 86\% on AgentTrap (Figure~\ref{fig:sources}).

We use the reported probability for every judge decision. Grok's boolean verdict disagrees with that probability on 242 of 1,947 MSD records and 253 of 1,384 SD records. GPT-5.4 mini has 10 such disagreements.

\section{Reproducibility}
\label{app:repro}

In the benchmark artifacts, SkillScreen is named \emph{Direct} to distinguish it from the scanner-context variants of Section~\ref{sec:context}. Run configurations pin the data, model, code, adapter, calibration and threshold. Before scoring test records, evaluation jobs replay validation examples to check the export. The code, model and per-skill predictions are available from the project page (\url{https://jerroldh.github.io/SkillScreen-webpage/}). The reproduction package recomputes Table~\ref{tab:main} offline from 66 prediction files (90,111 rows) covering its methods on MSD and SD, and includes the run configurations, split manifests, benchmark metadata and the data-preparation, training and scoring scripts. It bundles no skill text or API responses; a script restores the benchmark text from the pinned MaliciousSkillBench release.

\paragraph{Deployment releases.}
We release \href{https://huggingface.co/datasets/SkillScreen/SkillScreen-Data}{SkillScreen-Data}, a community resource containing 13,514 public-source skills with upstream labels, including 11,831 complete packages, alongside \href{https://huggingface.co/SkillScreen/SkillScreen-2B-Markdown/tree/v0.1.0}{8K Markdown} and \href{https://huggingface.co/SkillScreen/SkillScreen-2B-Package/tree/v0.1.0}{32K package models}.

We select thresholds from three-fold out-of-fold (OOF) predictions, keeping normalized Markdown duplicates within folds, then train each final model on its full eligible pool. At thresholds targeting 99\% OOF recall, the corresponding OOF FPR is 15.71\% for Markdown and 0.35\% for package inputs on their respective data pools. Model cards provide the thresholds, predictions and training details.

\end{document}